\documentclass[preprint,12pt,authoryear]{elsarticle}

\usepackage[hyphens]{url}
\usepackage{xcolor}
\usepackage{amsmath} 
\usepackage{amssymb}  
\usepackage{amsfonts}
\usepackage{tikz}
\tikzstyle{mybox} = [draw=black, very thick, rectangle, rounded corners, inner ysep=5pt, inner xsep=5pt]

\begin{document}

\begin{frontmatter}


\title{COVID-19 epidemic control using\\ short-term lockdowns for collective gain}
\author{Mauro Bisiacco}
\address{Department of Information  Engineering, University of Padova, Padova, Italy (e-mail: bisiacco@dei.unipd.it)}
\author{Gianluigi Pillonetto}
\address{Department of Information  Engineering, University of Padova, Padova, Italy (e-mail: giapi@dei.unipd.it)}

\maketitle
\begin{abstract}
While many efforts are currently devoted to vaccines development and administration,
social distancing measures, including severe restrictions such as lockdowns, remain fundamental tools to contain the spread of COVID-19.
A crucial point for any government is to understand, on the basis of the epidemic curve, 
the right temporal instant to set up a lockdown and then to remove it.
Different strategies are being adopted with distinct shades of intensity. 
USA and Europe tend to introduce
restrictions of considerable temporal length. 
They vary in time: a severe lockdown may be reached and then gradually relaxed.
An interesting alternative is the Australian model
where short and sharp responses have repeatedly tackled 
the virus and allowed people a return to near normalcy. 
After a few positive cases are detected, a lockdown is immediately set.
In this paper we show that the Australian model can be generalized and given 
a rigorous mathematical analysis, casting strategies of the type
\emph{short-term pain for collective gain} in the context of \emph{sliding-mode control}, an important branch of nonlinear control theory.
This allows us to gain important insights regarding how
to implement short-term lockdowns, obtaining a better understanding of
their merits and possible limitations.
Effects of vaccines administration 
in improving the control law's effectiveness are also illustrated.
Our model predicts the duration of the severe lockdown 
to be set to maintain e.g. the number of people in intensive care under a certain threshold.
After tuning our strategy exploiting 
data collected in Italy, it turns out that COVID-19 epidemic 
could be e.g. controlled 
by alternating one or two weeks of complete lockdown with one or two months of freedom, respectively.
Control strategies of this kind, where the lockdown's duration is well circumscribed, 
could be important also to alleviate coronavirus
impact on economy. 
\end{abstract}

\end{frontmatter}

\section{Introduction}
Important containment measures were implemented in Wuhan (China)
to  limit the diffusion of SARS-CoV-2 \citep{Fei2020,Wu2020,Guan2020}. 
Despite these efforts, such new strain of coronavirus has spread all over the world \citep{Velavan2020,Wittkowski2020}.
The development and administration of vaccines is now playing a fundamental role to face the epidemic. 
However, 
the use of masks and the
introductions of social distancing measures, including 
severe restrictions like lockdowns, can still play an  important role in the future to control the epidemic spread,
possibly also to deal with virus variants \citep{Crisanti2020,VirusVar2021}.\\
In this scenario, any government is called to understand, on the basis of the epidemic curve, 
the right temporal instant to set up a lockdown and then to remove it.
Different strategies are being adopted with distinct shades of intensity. 
USA and many countries in Europe tend to introduce
restrictions of considerable temporal length. 
They change as time progresses: a severe lockdown may be reached and then gradually relaxed.
An interesting alternative is given by the Australian model
where short and sharp responses have repeatedly tackled 
the virus and allowed people to almost return to a normal life \citep{WAtoday2021}. 
After a few positive cases are detected, a lockdown is immediately set.
As an example, on January 2021 in Perth, Australia, a complete lockdown 
immediately started on Sunday just after assessing that one hotel quarantine worker tested positive \citep{TNYT2021}. 
Hence, suddenly, two million people were forced to stay at home for a few days. 
Similar facts happened in Brisbane and Sydney.
As said, so fast responses have not been implemented in USA or Europe
where people follow longer lockdowns of time-varying intensity, typically
including less severe constraints.\\ 
 
 In this paper we show that the Australian model can be formalized, generalized and  
studied in mathematical terms. The effects of vaccines administration in improving the control law's effectiveness will be also illustrated. To obtain this,  we use dynamic systems models to describe  
SARS-CoV-2 spread and cast strategies of the type
\emph{short-term pain for collective gain} in the context of \emph{sliding-mode control}, an important branch of nonlinear control theory \citep{Sliding2012,SlidingBook,Ibeas2013}.
Our analysis provides important insights as how
to implement lockdowns of brief duration to control SARS-CoV-2 dynamics, a problem also
discussed e.g. in \citep{Sadeghi2021,Borri2021,Garetto2021}, 
leading to further insights on 
their merits and possible limitations. The  health system's target 
should be to maintain the number of people infected or in intensive care below a certain level,
see also recent works on control of SARS-CoV-2 like \citep{CCGondim2020,CCTsay2020,CCKohler2020,CCBerger2020}.
When such threshold is exceeded, 
our control rule permits to come back to the right value
exponentially fast and the duration of the severe lockdown which has to be set to reach the goal can be predicted.
Interesting scenarios emerge, 
in line with the Australian strategy.
For instance, by tuning the model and the predictive rule 
using data collected in Italy, it turns out that COVID-19 epidemic 
could be controlled, maintaining the number of people in intensive care under the critical threshold,
by alternating for instance one or two weeks of complete lockdown
(allowing e.g. only essential shopping and medical appointments)
with, respectively, one or two months of freedom (where e.g. people can move everywhere - except for traveling to/from other countries - just using masks and respecting basic measures of social distancing).
Control strategies of this kind, relying on lockdowns of well circumscribed duration, 
could be important also to alleviate coronavirus impact on economy.\\

%
%
%

In order to describe COVID-19 dynamics, many models have recently appeared.
Examples are  \emph{network models} where a network of interacting individuals
describes the population.
This requires to introduce many unknown parameters
whose determination could be difficult from the available measurements \citep{Keeling2005,Pastor2015,Pellis2015,Pastore2014}.  
\emph{Compartmental models} are a simpler description
with the population assumed to be
well-mixed and split into different categories. The most popular model inside this class 
is the SIR model which includes three compartments containing susceptible (S), infected (I) and removed (R) individuals \citep{Kermack1927} 
and variants like those described in \citep{Bootsma2007,Capasso1978,Korobeinikov2005,Liu1987}.
For what concerns COVID-19 pandemic, many SIR-type models 
have been also developed e.g. to describe how people can react to knowledge of infections and 
risk of death \citep{Casella2020,Lin2020,Anastassopoulou2020,Weitz2020}. 
Other work has been addressed also to the so-called {\sl spatially explicit models} where homogeneity assumption is mitigated
by describing the spatial infection along both time and space
 \citep{Gatto2014,Riley2015,Bertuzzo2010,Mari2019}. 
 See also \citep{OSullivan2020} and \citep{Gatto2020} 
where mobility data are also exploited
to infer the contact rate's variations regulating  COVID-19 spread.\\

Potentials of sliding mode in controlling COVID-19 has been recently discussed in
\citep{Nunez2021} 
using the SEIR model. In comparison with that work, our control strategy 
is different and exploits the so called \emph{chattering phenomenon} to alternate
freedom and lockdown periods of suitable length.
In addition, a theoretical analysis regarding the convergence of 
our control scheme is included by adopting a variety of epidemiological models. 
We will consider the SEIR and SAIR models which include, respectively, the class of exposed (E), 
who are host for infectious but cannot yet transmit the disease, and the class of asymptomatic people,
who may play an important role in transmitting COVID-19 \citep{Wang2020,Crisanti2020}.
This will permit to maintain the exposition simple yet using models
well suited to describe COVID-19 dynamics by tuning it with prior information from the literature and 
some epidemiological data collected in Italy.
Notably, it will turn out that the proposed control algorithm is robust, 
in the sense that it requires only a coarse knowledge of SEIR or SAIR parameters.
Then, in Appendix we will further extend the theory by discussing an even more complex model,
the SEAIR \citep{Sharifi2017},  
that includes exposed people as well as asymptomatic/symptomatic infected.\\
The paper is organized as follows. 
Section \ref{Models} describes the models used to describe 
SARS-CoV-2 dynamics and the control law.
In Section \ref{Results}, first the main theoretical findings obtained in this paper are reported
and then the control strategy is tested on realistic case studies.
Conclusions then end the paper while the Appendix contains the proofs
of the mathematical results cited in the paper.

\section{SARS-CoV-2 modeling and control}\label{Models}

We now introduce the models used to simulate the epidemic and our nonlinear control law.
Let us consider four classes $S(t),I(t),E(t)$ and $R(t)$ which evolve in time and 
contain, respectively, susceptible, infected, exposed and removed 
people. They are normalized, hence their sum is equal to one for any temporal instant $t$.\\
The following set of differential equations then define the \emph{SEIR model} 
 \begin{subequations}\label{SEIReq}
\begin{align}
\dot{S}(t)&= - \beta(t)S(t)I(t) \\
\dot{E}(t)&= \beta(t)S(t) I(t) - \epsilon E(t) \\ 
\dot{I}(t)&= \epsilon E(t)- \gamma I(t) \\ 
\dot{R}(t)&= \gamma I(t)
\end{align}
\end{subequations}
where, scrolling through the equations from the bottom, 
the scalar $\gamma$ regulates the rate with which infected people heal or die,
$\epsilon$ indicates the rate with which exposed become infected, 
while $\beta(t)$ is the infection/contact rate that 
describes the time-varying interaction between susceptible and infected.
It is also useful to recall the definition of \emph{reproduction number}
\citep{Diekmann1990,Wallinga2004,MultiRN}.
When $S(t) \approx 1$ 
and assuming constant $\beta(t)=\beta$,
it is given by 
 \begin{equation}\label{RN}
\text{RN} = \frac{\beta}{\gamma},
\end{equation}
and RN values smaller than one indicate that the number of infected  is exponentially decreasing.\\

Using instead the \emph{SAIR}, infected are divided into two classes. 
 The first class contains asymptomatic/paucisymptomatic $A(t)$ who 
 can directly recover with a rate established by $\epsilon_2$.
 The other ones move to the second class of infected $I(t)$ with a rate $\epsilon_1$ and then recover
 with a rate $\gamma$.  One thus has
 \begin{subequations}\label{SAIReqMT}
\begin{align}
\dot{S}(t)&=- \beta(t)S(t)\big(A(t)+I(t)\big) \\
\dot{A}(t)&= \beta(t)S(t)\big(A(t)+I(t)\big) - \big(\epsilon_1+\epsilon_2\big)A(t) \\
\dot{I}(t)&= \epsilon_1A(t)-\gamma I(t) \\ 
\dot{R}(t)&=\epsilon_2A(t)+\gamma I(t) 
\end{align}
\end{subequations}
where $\beta(t)$ now describes the temporal evolution of the interaction between susceptible $S(t)$
and the two classes $A(t),I(t)$ of infected. 
Following \citep{Diekmann1990,MultiRN},
a reproduction number can be defined  
even for this kind of model with multiple classes 
by computing the largest modulus of the eigenvalues of the so-called 
next-generation matrix.

The time-varying rate $\beta(t)$ is the result not only of biological but also of social factors. 
Hence, its value, as well as that of RN, contains also indication on the level of freedom and sociality
of a population. In our mathematical description, it represents the unique parameter which can be modified
by setting some restrictions. We will assume that it can take two values: 
$\beta_F$ in absence of restrictions (freedom) and $\beta_L$ during the lockdown.
We also use $I_0$ to denote the desired equilibrium point, i.e. 
the target value for the number of infected $I(t)$.
Ideally, one would like $I_0=0$ but this is not possible since, in absence of vaccines,
it would require a lockdown of infinite temporal length. Our aim is instead 
to design a control law that maintains the system
around a small value of $I_0$, so that the number of people in intensive care can be managed by the health system, 
by mimicking the Australian strategy. Our approach is then formalized below.
In its description, $\lambda>0$ 
and $\phi \geq 0$ are two design scalar parameters
whose role to control the epidemic will be discussed in depth later on.
\medskip

\hspace*{-1mm}\begin{tikzpicture}
\node [mybox] (box){%
    \begin{minipage}{.96\textwidth} 
{\bf{\large Control law}}:\\
\begin{itemize}
\item Assume that people are in a situation of full freedom, i.e. there are no restrictions at instant $t$.
Then, set the lockdown if
$$
\lambda \big(I(t)-I_0\big) + \dot{I}(t) >\phi \geq 0. 
$$ 
\item   Assume that people are under restrictions, i.e. the lockdown holds at instant $t$.
Then, remove all the restrictions if
$$
\lambda \big(I(t)-I_0\big) + \dot{I}(t) < -\phi \leq 0. 
$$ 
\end{itemize}
    \end{minipage}
};
\end{tikzpicture}%
\medskip

The control law proposed above relies on a sophisticated nonlinear control technique known as \emph{sliding-mode control}, see also Appendix for details. Such an approach tries to bring (and maintain) the state trajectory on a given sliding surface, where the dynamics becomes completely independent of the model parameters and can be tuned accordingly to a desired target
(this point is further described in second remark present at the end of Section \ref{Sec1A} in Appendix). 
In our case, the sliding surfaces for the SEIR and the SAIR models satisfy, respectively, the following equations 
\begin{subequations}\label{SlidingSurface}
\begin{align}
\epsilon E +(\lambda-\gamma) I &= I_0 \lambda \quad \text{(SEIR)}\\
 \epsilon_1 A+(\lambda-\gamma) I &= I_0 \lambda \quad \text{(SAIR)}. 
 \end{align}
\end{subequations}
For what regards the control parameters, the scalar $\lambda$ establishes how aggressive the control is. Large values of $\lambda$
make the system approach faster to the sliding line. 
In addition, we also want to suitably alternate the  
periods of lockdown and freedom.
This can be obtained by tuning the parameter $\phi$. 
The choice $\phi>0$ is needed in practice since it induces chattering, i.e. an oscillation around $I_0$.
This is needed for avoiding an infinite frequency switching between freedom and lockdown: in any real-world situation, at least some days are required for changing the level of the restrictions. In practice, in our simulations it will be set to a 
conveniently low-value
in order to obtain a suitable switching period.
 In its ideal version without such threshold, i.e. with $\phi=0$, our control law activates a perfect sliding-mode 
 whose functioning 
 also requires some quite mild assumptions to be satisfied, 
 as detailed in Appendix.\\
 
In what follows, theoretical results concerning the proposed algorithm will be first illustrated.
Then, we will apply it to simulate the control of SARS-CoV-2 epidemic.
Notably, we will see that the same law here described can be applied to either SEIR or SAIR 
guaranteeing the same control properties. 
Further generalizations will be studied in Appendix
by introducing the SEAIR model, where 
the classes of exposed, asymptomatic and symptomatic infected are all included.

\section{Results}\label{Results}
\subsection{Main theoretical findings} 
We summarize the two main theoretical results (described in detail in Appendix)
regarding the proposed control law.

\begin{itemize}
\item there always exist values of $\lambda$, known from a vague knowledge of SEIR or SAIR parameters,
that ensure perfect convergence to the desired target $I_0$ when $\phi=0$. Together with $\lambda$, the value of $\phi$
can then be adjusted to regulate the chattering phenomenon; 
\item the epidemic control is robust since global convergence holds, i.e. the value $I_0$ is reached starting from any initial condition of  either the SEIR or the SAIR model (at least if $\lambda>0$ is small enough). In addition, the control is guaranteed also when system parameters are known with scarce precision.
\end{itemize}

\subsection{Tuning the model and the control law} 

\begin{figure*}
	\begin{center}
		\begin{tabular}{c}
			\hspace{-.2in}
			{ \includegraphics[scale=0.24]{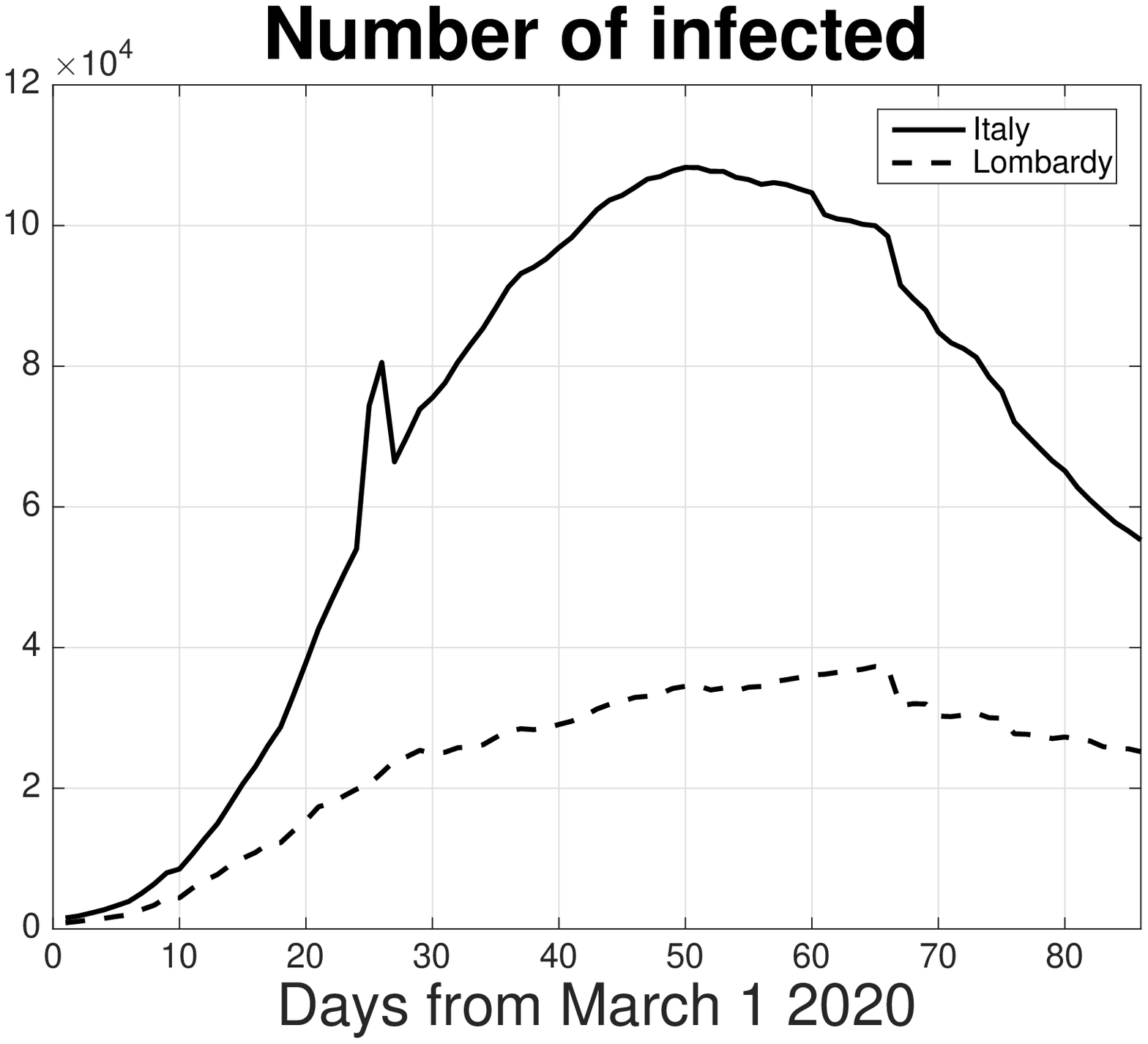}} \
			{ \includegraphics[scale=0.24]{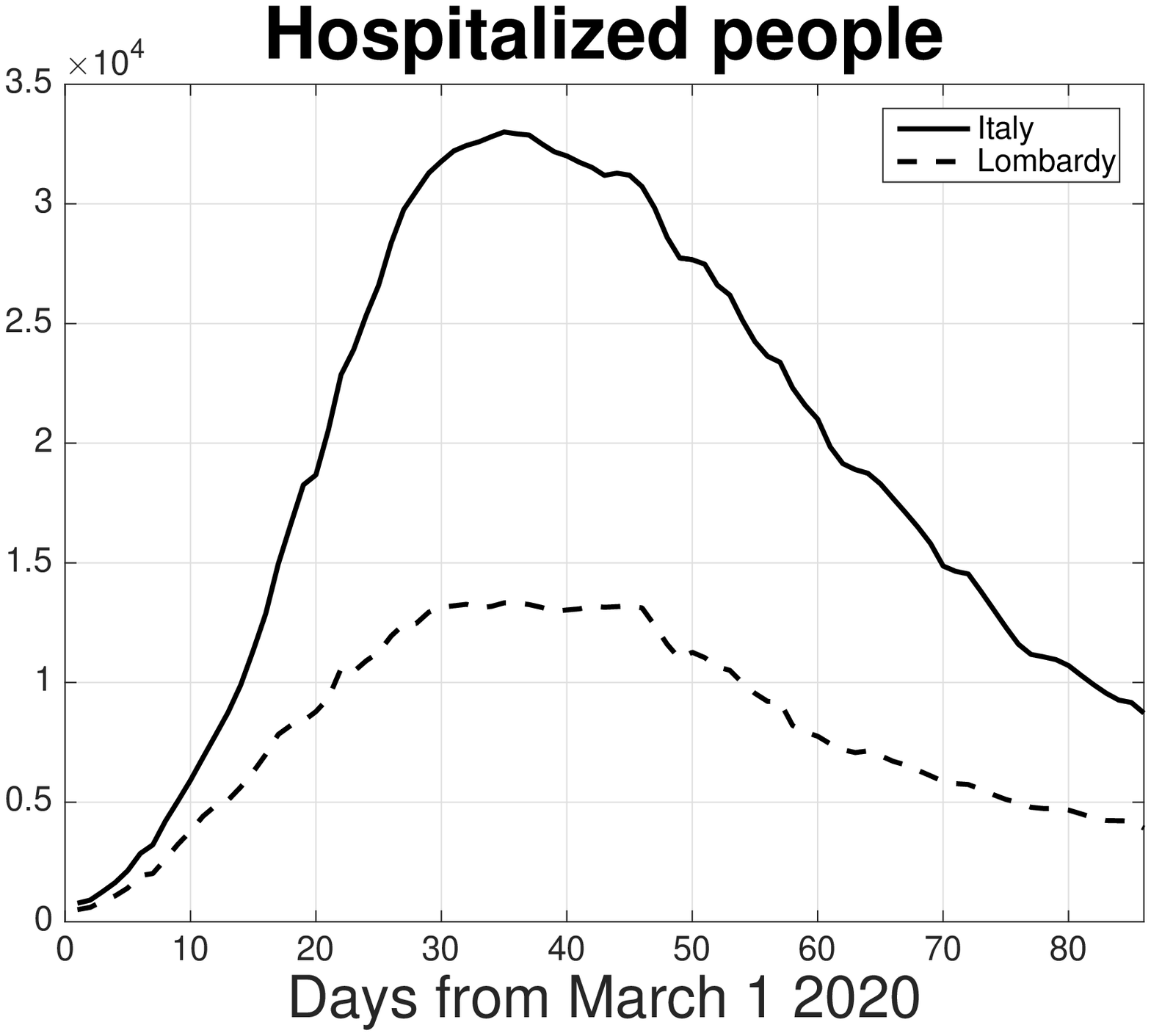}} \
			{ \includegraphics[scale=0.24]{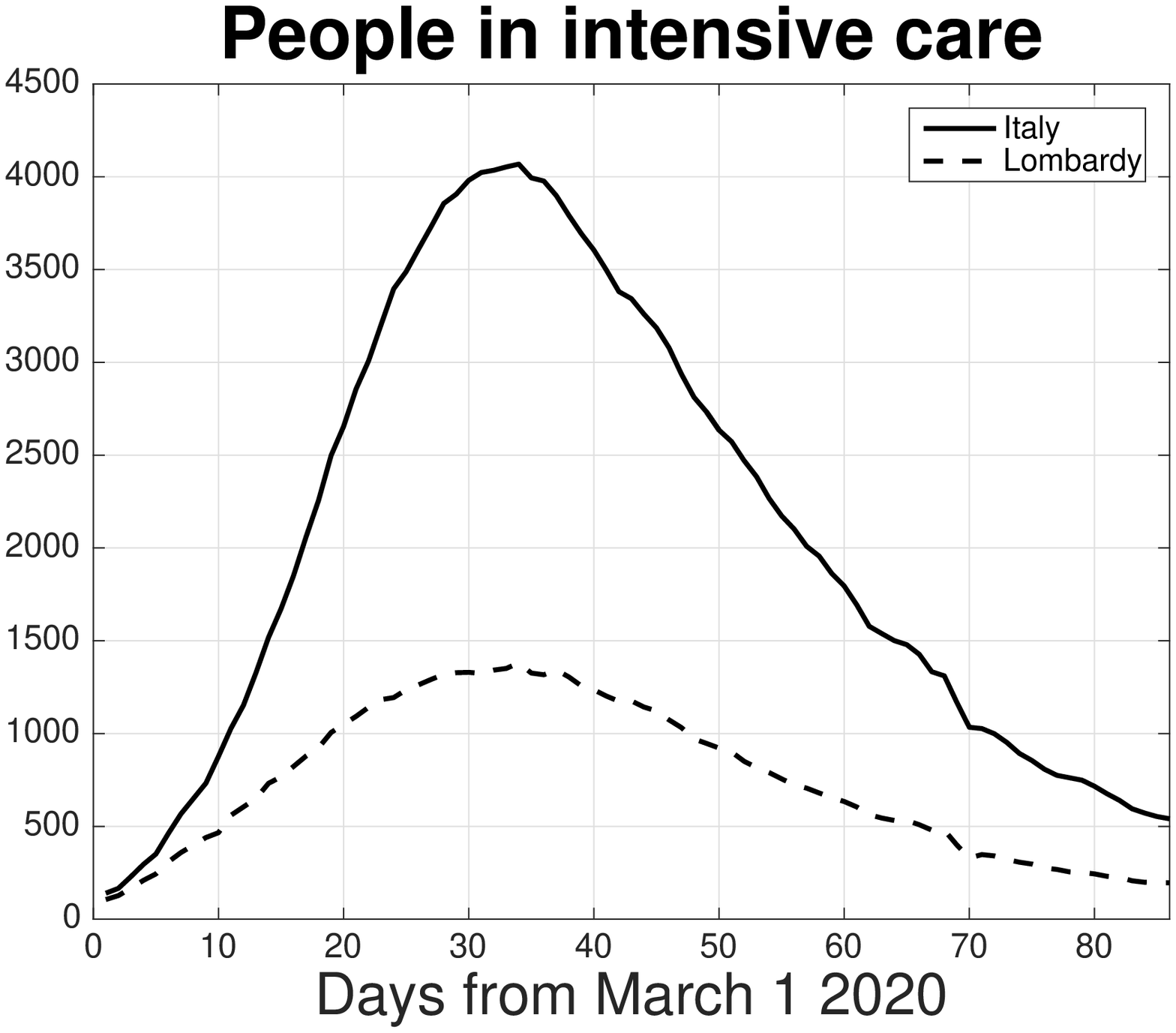}} 
		\end{tabular}
		\caption{Number of diagnosed infected (left), hospitalized people (middle) and people in critical care (right)
		collected on a daily basis in Italy (solid line) and Lombardy (dashed) during the first wave of infection. 
		The first instant corresponds to March 1,  2020.
		The first lockdown started on March 9, 2020, and the temporal interval of the figure then extends until its end.} \label{Fig1}
	\end{center}
\end{figure*}

In what follows, due to space constraints, we focus on SEIR
 simulation and control. Results returned by the SAIR
 lead to the same conclusions illustrated below
and can be found in Appendix.\\
SEIR parameters are defined by exploiting available 
information on COVID-19 dynamics, e.g. \citep{Crisanti2020,Gatto2020,Giordano2020}.
All the states are normalized, so that $S(t)+E(t)+I(t)+R(t)=1 \ \forall t$.
The nominal value for $\gamma$ is set to $0.05$ 
which means that people heal on average in 20 days.
To test the robustness of our control we will also make 
such rate vary in the interval $[0.03,0.07]$ (from two weeks to one month to heal).
The nominal value for $\epsilon$ is $0.2$ 
which means that people pass from exposed to infected in almost 5 days.
The uncertainty interval for the incubation period is then $[0.1,0.3]$ (from three to 10 days) \citep{Incubation}.
To have insights on the value $\beta_F$ that the contact rate $\beta(t)$
may assume during the freedom period, we use the reproduction numbers
calculated by the Italian Ministry of Health during August, September and October, 2020 \citep{SaluteSite}.
These months are especially significant since  freedom of people  
much increased due to holiday relax and then also to the return to 
work and school. By monitoring the weekly reports present in \citep{SaluteSite},
the mean of the $95\%$ upper bounds of the reproduction number RN
turns out to be 1.3. This then defines $\beta_F$ through 
\eqref{RN}.  We will also investigate the use of different reproduction numbers 
falling in the interval
$[1.2,1.6]$. Finally, to simulate a severe lockdown period, 
the contact rate $\beta_L$ during the restrictions is set to $0.01$
with the uncertainty interval extending up to the value $0.02$.
This reflects the RN values measured on April 2020 during the lockdown in Italy, around 0.2-0.5 \citep{SaluteSite}.
To simulate the epidemic, the initial value of the infected and exposed is $0.1\%$ of the population.\\
For what regards the control law, 
the target is to maintain the value of $I(t)$
to $0.2\%$ of the population. For this purpose, we will set $\phi=10^{-4}$
and we will adopt two values for $\lambda$, equal to $0.2$ and $0.6$, respectively.\\
To simulate the control of the SARS-CoV-2 spread, 
knowledge of $I(t)$ and of its first-order derivative is assumed. 
Since  the true number of infected is not perfectly known during an epidemic,
in practice $I(t)$ can be inferred from epidemiological measurements, 
like the number of hospitalized or diagnosed infected 
reported in Fig. \ref{Fig1}. They are collected  on a daily basis in Italy
and Lombardy (the Italian region where most of the outbreak happened)  
during the first wave of infection, starting from March 1, 2020. 
Continuous-time profiles could be obtained by fitting 
such data e.g. through cubic smoothing splines \citep{Wahba:90}.
Note that, when using the SAIR model, estimates of $I(t)$
can be even more reliable since they do not need to include asymptomatic people who are
in the most difficult class to be detected.
 The number 
 of patients in critical care $I_c(t)$ can 
 be also much informative about COVID-19 dynamics 
and one strategy adopted in \citep{pillonetto2020tracking} to estimate the number of infected is 
to assume it proportional to $I(t)$, i.e. $I(t)=H I_c(t)$. 
For future illustration purposes, it is useful to gain some information on this scale factor.
In Italy, a robust lower bound for $H$ appears to be $50$, a value 
that follows from simple considerations on the time-courses of diagnosed infected
and people in ICUs (intensive care units) 
displayed in Fig. \ref{Fig1}.
Since the Italian population is around $60$ million, 
our control target $I_0$ then corresponds to maintaining the number of people
in intensive care around 2400. This number is smaller than the critical threshold 
estimated in Italy as the $35\%$
of the (almost) 10000 places available in ICUs. 

%
\begin{figure*}
\center
{\includegraphics[scale=0.4]{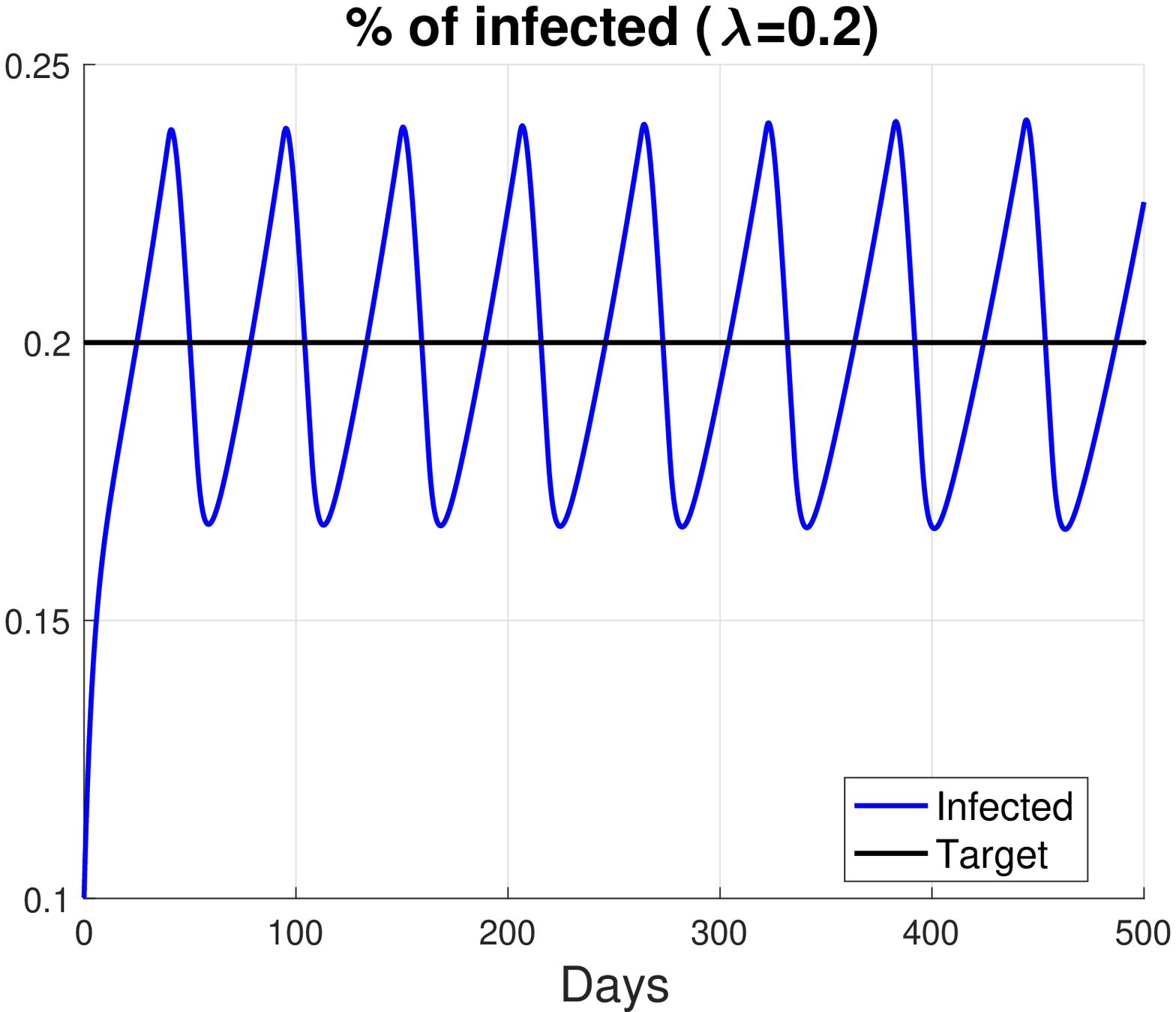}} \\ \vspace{0.2cm}
{\includegraphics[scale=0.4]{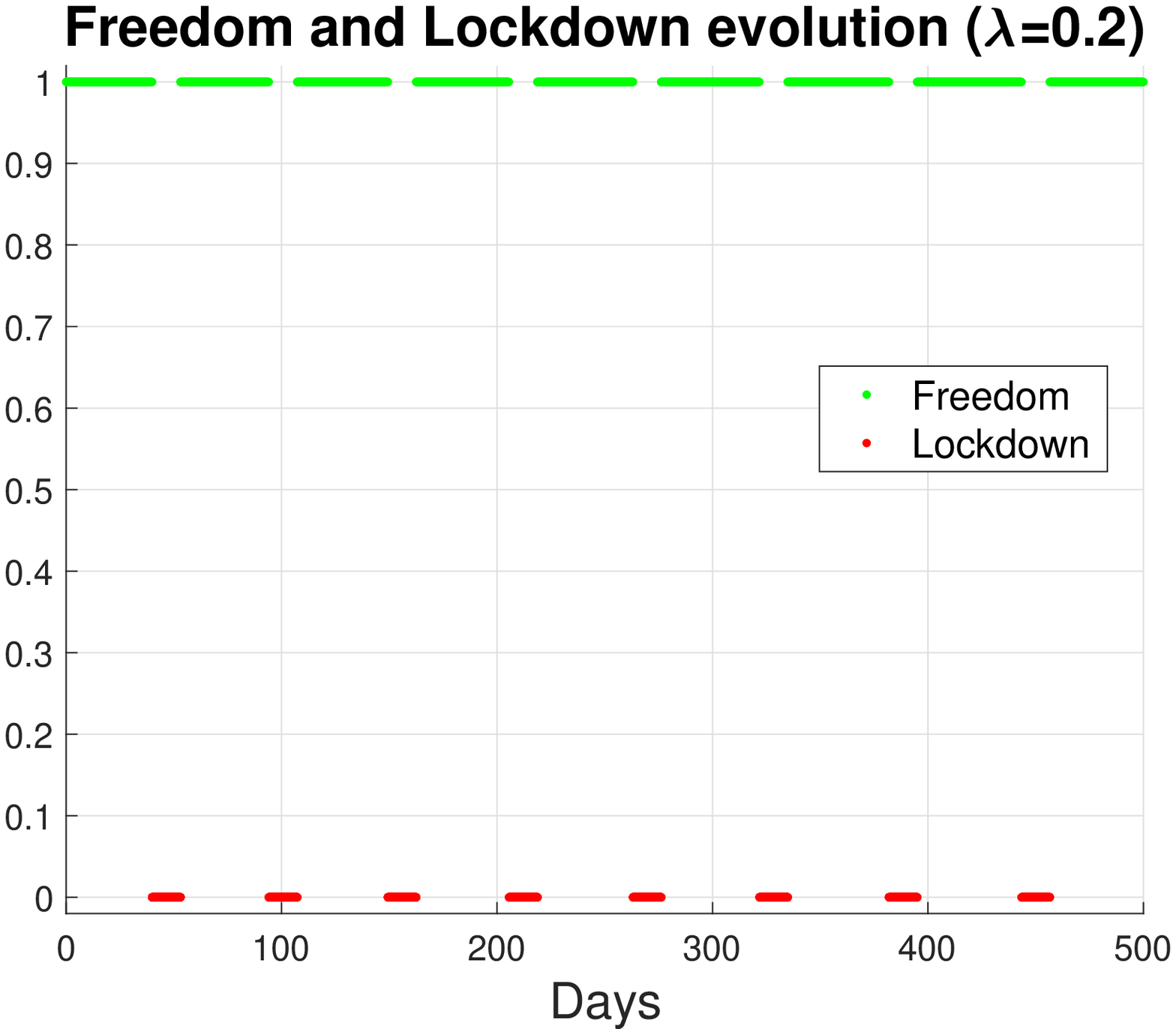}} \\ \vspace{0.2cm}
{\includegraphics[scale=0.4]{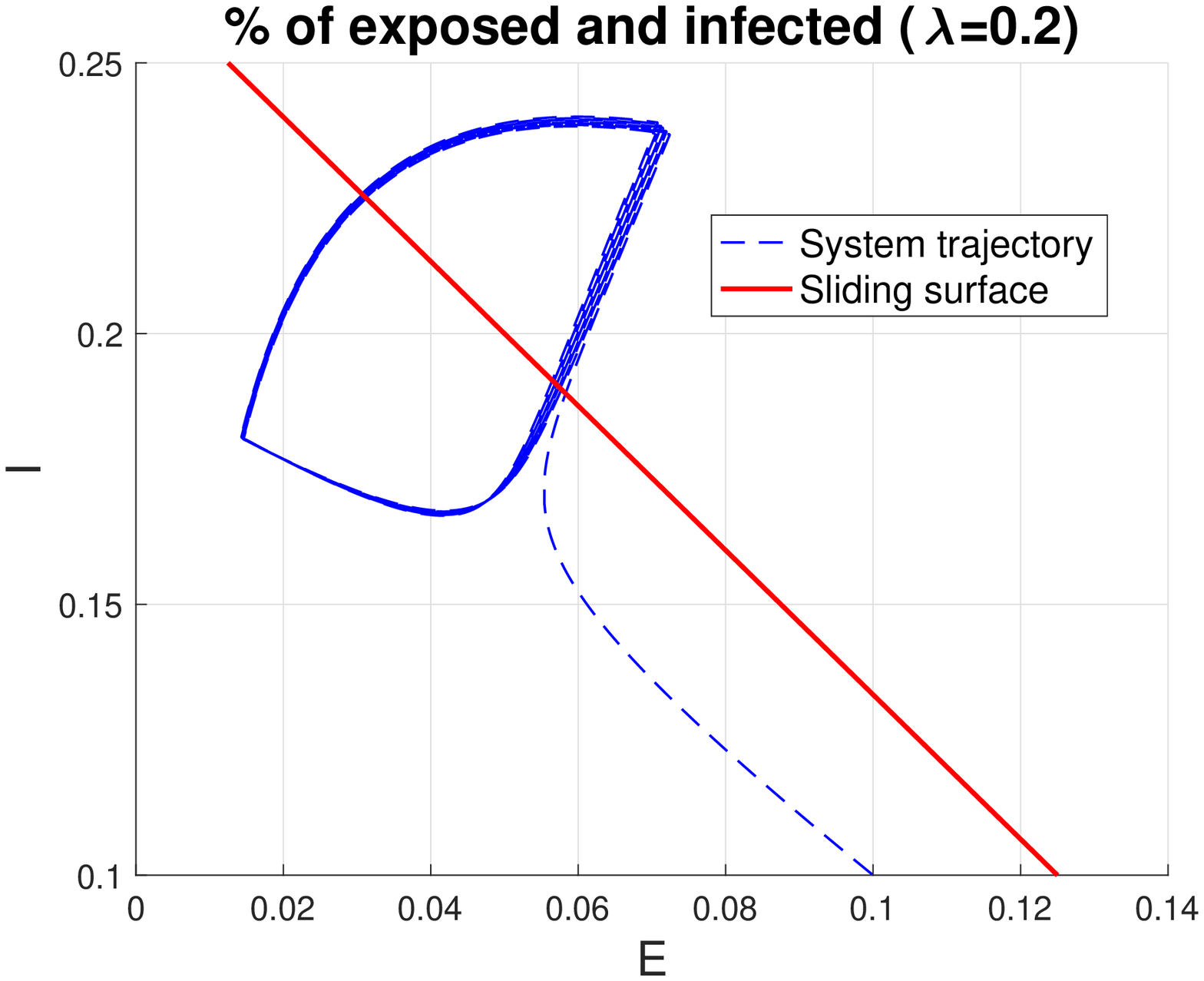}}
\caption{Results returned by the nonlinear control law with $\lambda=0.2$. \emph{Top} Percentage of infected (blue) and desired percentage (black) as a function of days.
\emph{Middle} Temporal evolution of freedom (green) and lockdown (red) periods. 
After a transient, almost two months of freedom and two weeks of severe lockdown are alternated.
\emph{Bottom} Trajectory describing how the percentage of exposed and infected varies in time (blue) and sliding surface (red).}
\label{Fig2}
\end{figure*}

\begin{figure*}
\center
{\includegraphics[scale=0.4]{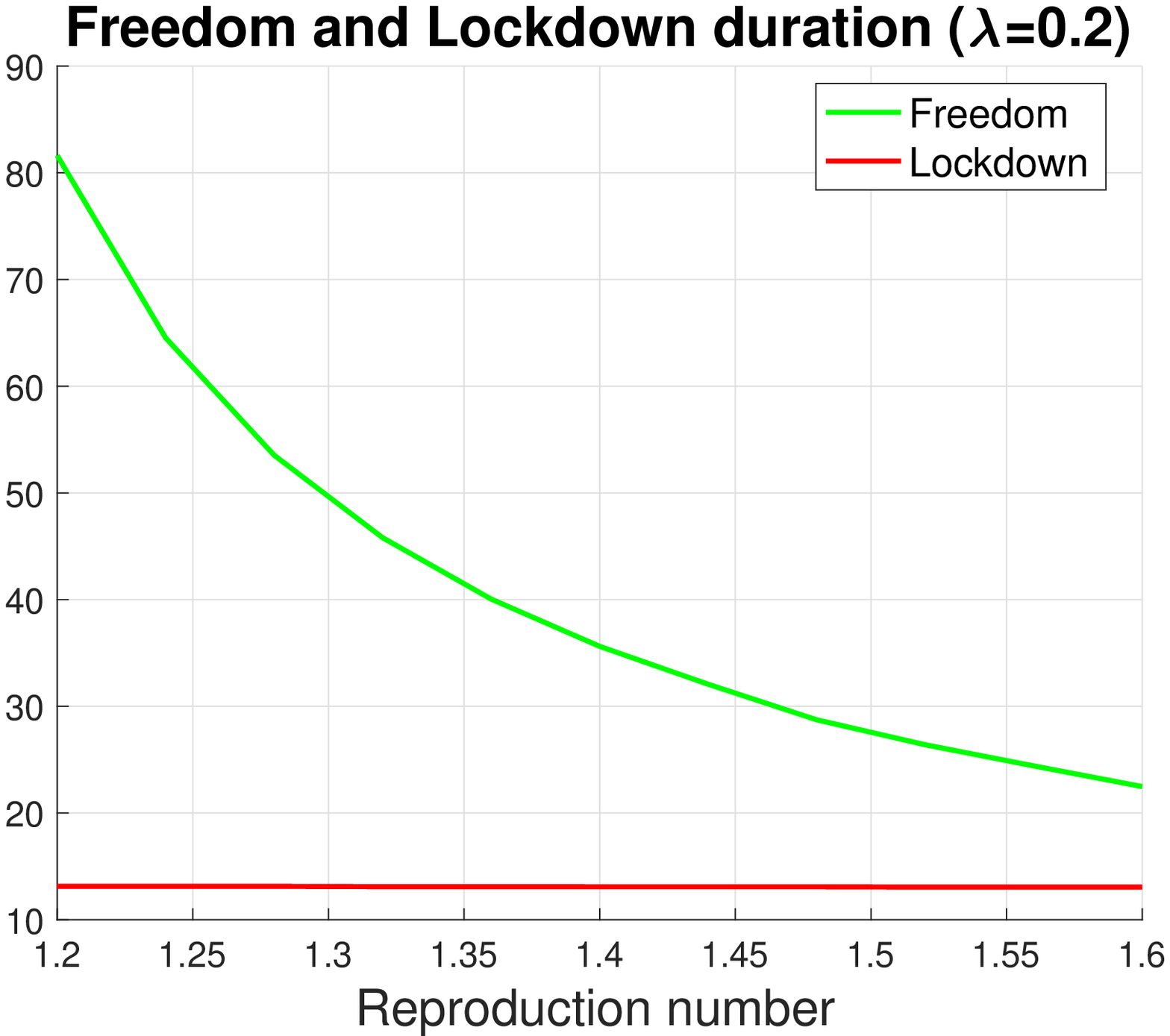}} 
\caption{Temporal duration of the freedom (green) and lockdown (red) period, as a function of the reproduction number  returned by the nonlinear control law with $\lambda=0.2$. 
The case with the reproduction number set to 1.3 corresponds to that detailed in Fig. \ref{Fig2}.}
\label{Fig3}
\end{figure*}

\begin{figure*}
\center
{\includegraphics[scale=0.4]{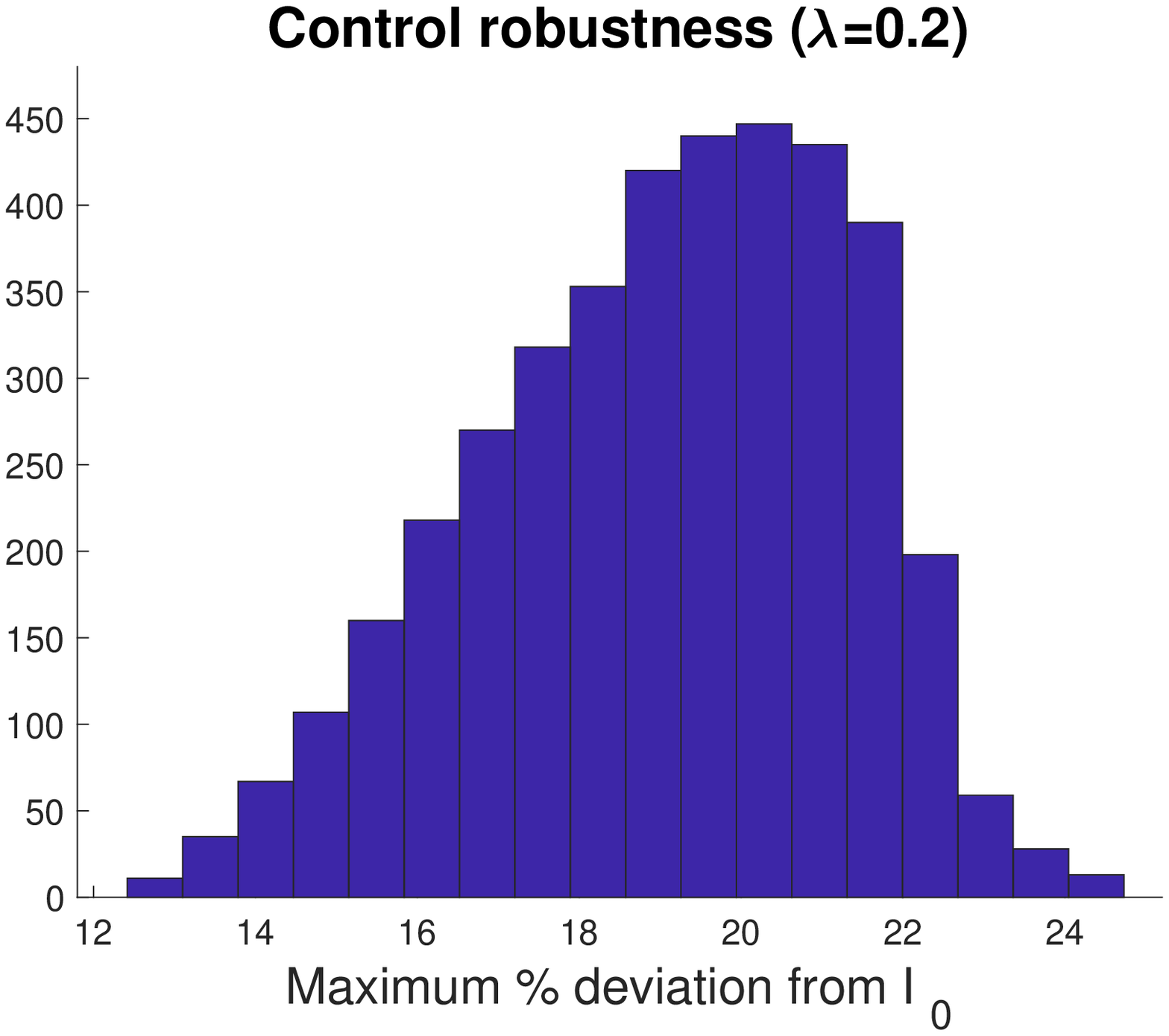}} 
\caption{Precisions in the control of the target value $I_0$ of infected using $\lambda=0.2$. The histogram contains the percentage deviations (computed after the initial transient) given by $100\% \max_t |I(t)-I_0|/I_0$ making SEIR parameters $\gamma$  and $\epsilon$ vary on the intervals $[0.03,0.07]$ and $[0.1,0.3]$, respectively.  The figure thus shows that in any case the target $I_0$ can be well controlled: the error never exceeds $25\%$.}
\label{Fig3b}
\end{figure*}

\begin{figure*}
\center
{\includegraphics[scale=0.4]{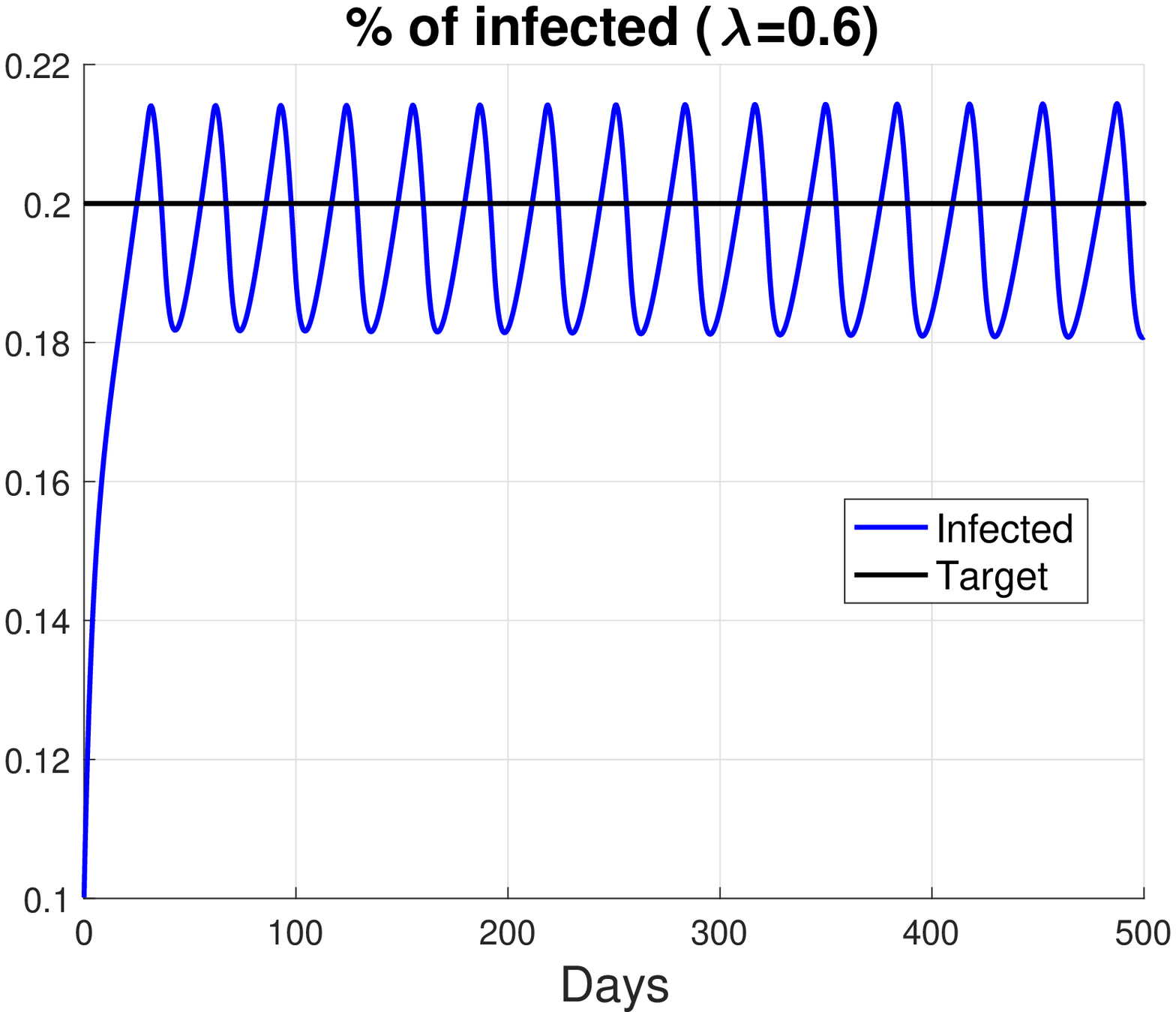}} \\ \vspace{0.2cm}
{\includegraphics[scale=0.4]{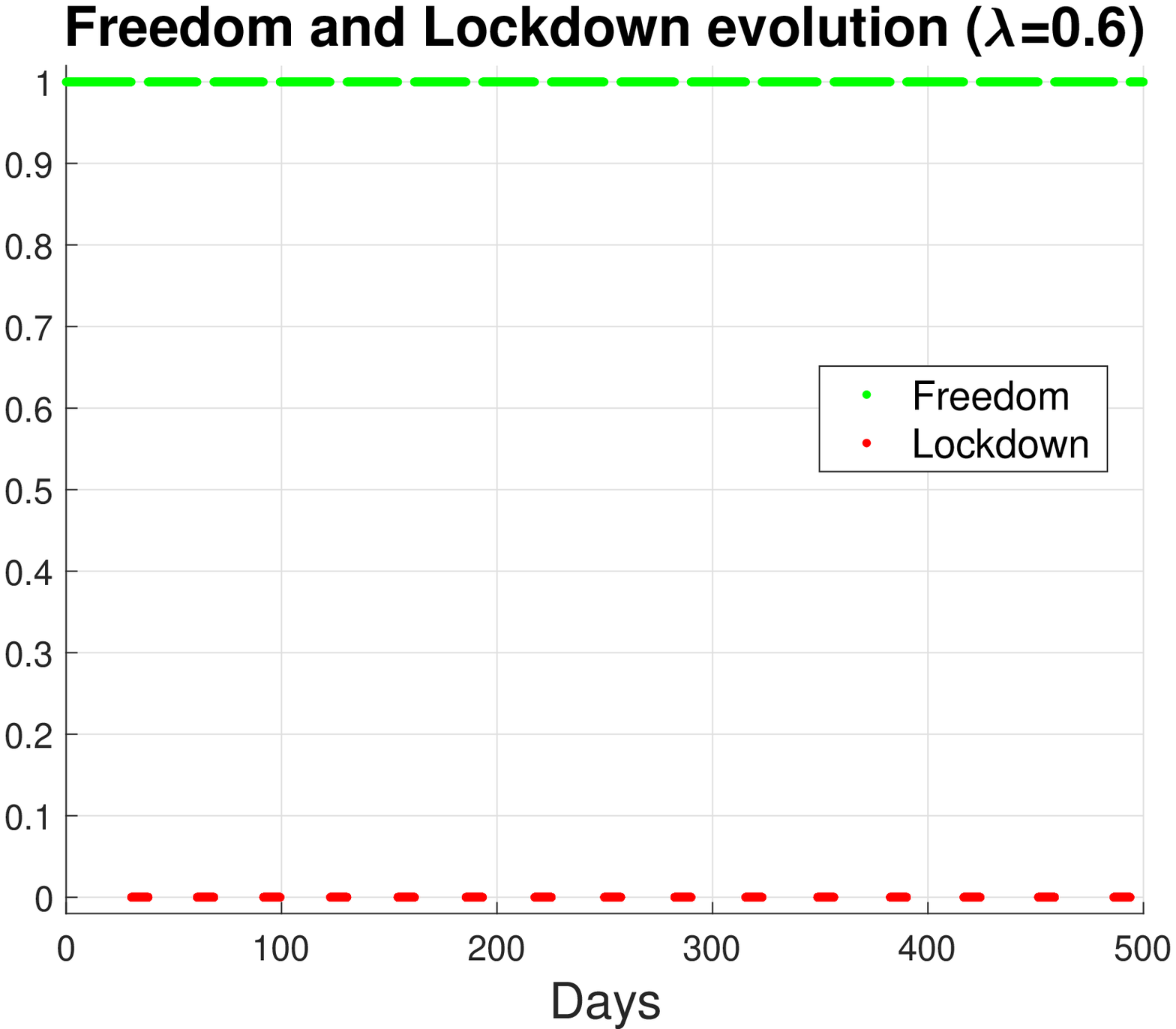}} \\ \vspace{0.2cm}
{\includegraphics[scale=0.4]{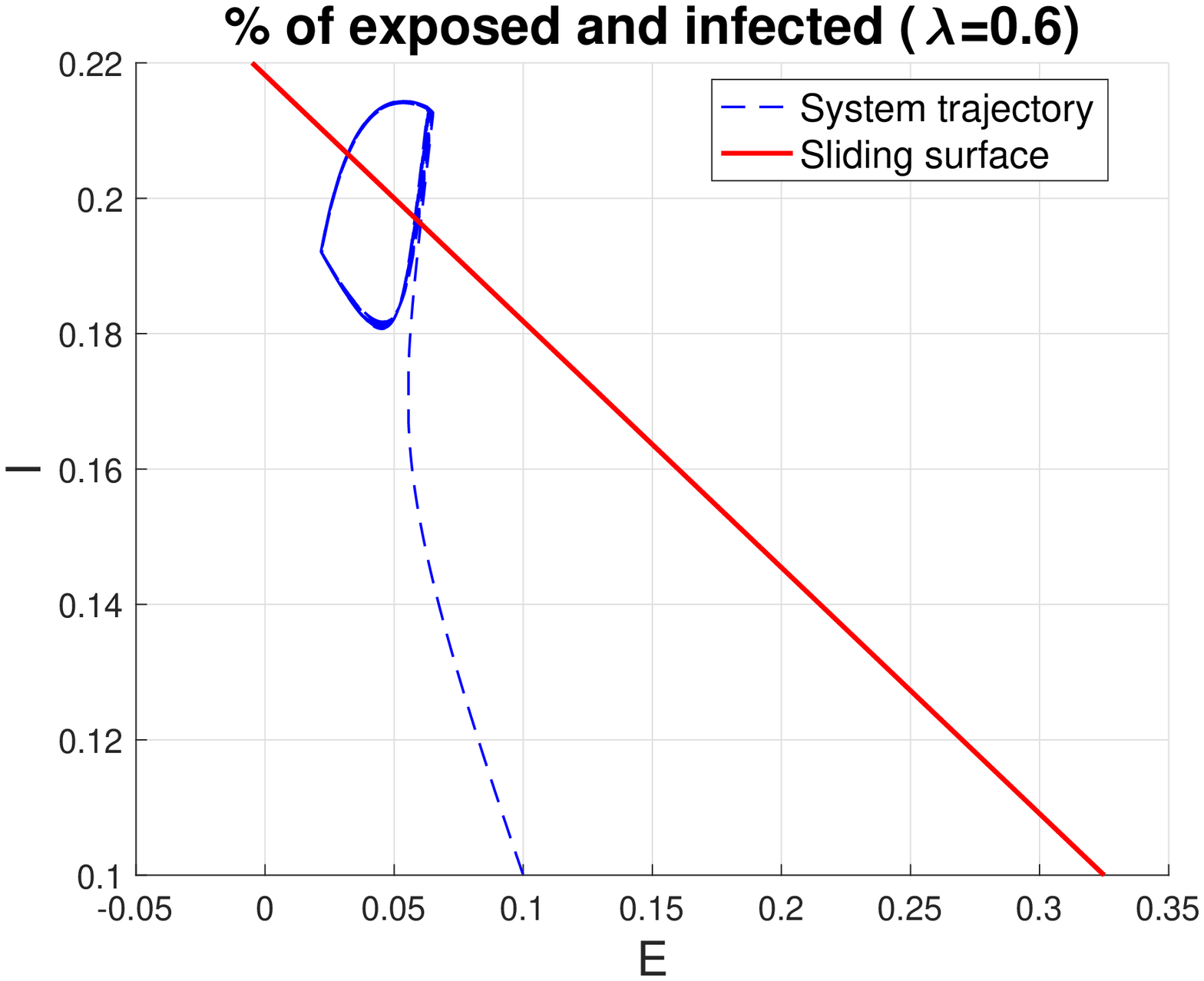}}
\caption{Results by the more aggressive control law with $\lambda=0.6$. \emph{Top} Percentage of infected (blue) and desired percentage (black) as a function of days. In comparison with the top panel of Fig. \ref{Fig2}
the amplitude of the oscillations of the infected is attenuated.
\emph{Middle} Temporal evolution of freedom (green) and lockdown (red) periods. 
After a transient, the control law now alternates almost one month of freedom and one week of severe lockdown.
\emph{Bottom} Trajectory describing how the percentage of exposed and infected varies in time (blue) and sliding surface (red).}
\label{Fig4}
\end{figure*}

\begin{figure*}
\center
{\includegraphics[scale=0.4]{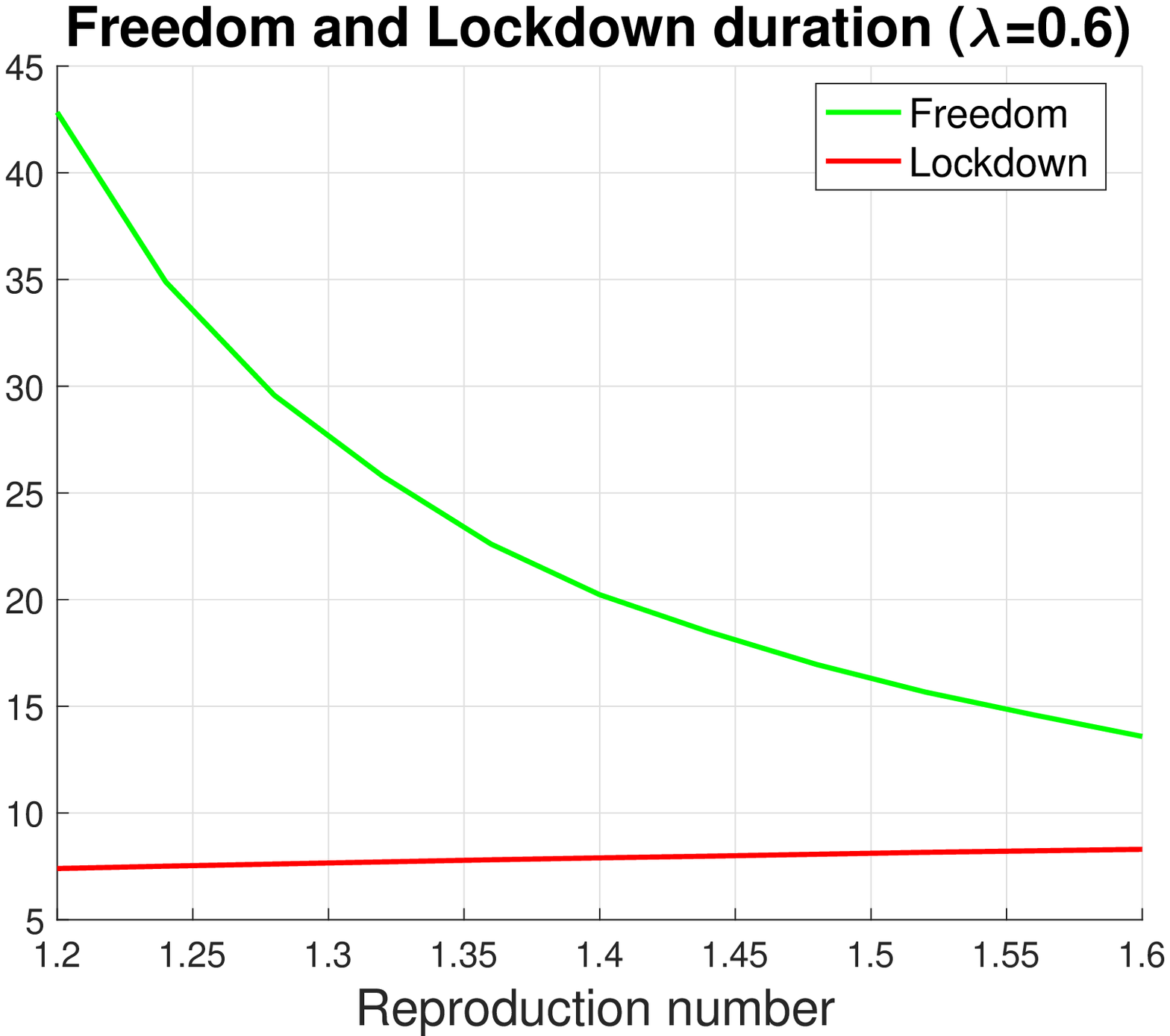}} 
\caption{Temporal duration of the freedom (green) and lockdown (red) period, as a function of the reproduction number  returned by the nonlinear control law with $\lambda=0.6$. 
The case with the reproduction number set to 1.3
corresponds to that detailed  in Fig. \ref{Fig4}.}
\label{Fig5}
\end{figure*}

\begin{figure*}
\center
{\includegraphics[scale=0.4]{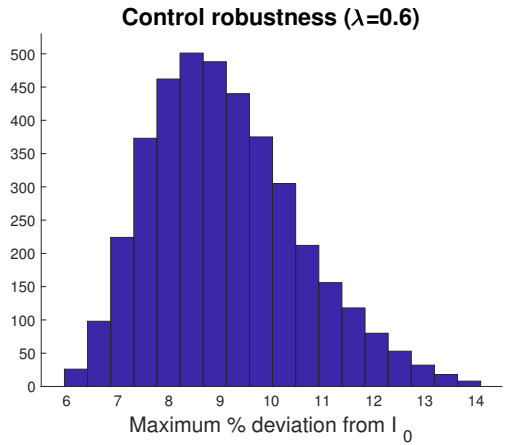}} 
\caption{Precisions in the control of the target value $I_0$ of infected using $\lambda=0.6$. The histogram contains the percentage deviations (computed after the initial transient) given by $100\% \max_t |I(t)-I_0|/I_0$ making SEIR parameters $\gamma$  and $\epsilon$ vary on the intervals $[0.03,0.07]$ and $[0.1,0.13]$, respectively. The figure thus shows that in any case the target $I_0$ can be well controlled: the error never exceeds $14\%$.}
\label{Fig5b}
\end{figure*}

\subsection{Control of SARS-CoV-2} 

First, we simulate the model using the nominal values for the SEIR parameters and setting $\lambda=0.2$
in the nonlinear control law. The top panel of Fig. \ref{Fig2} displays the percentage of infected (blue) and the 
target value (black). After a transient phase, one can see that $I(t)$ starts oscillating around the desired 
level $I_0$. The chattering is moderate, and also the infected peak (below $0.24\%$, an increment of the $20\%$ of $I_0=0.2\%$) does not lead
to exceed the critical threshold of people in intensive care.
The middle panel then shows how the control law, in a totally automatic way, 
alternates freedom (green)  and lockdown (red) periods. 
Interestingly, after getting rid of initial conditions effects, SARS-CoV-2 epidemic 
can be controlled using almost two months of freedom alternated with two weeks of severe restrictions.
This is quite in line with the Australian scenario which appears so confirmed 
through our model. Results are not so sensitive to the adopted values of the contact rate 
during the freedom period. Setting $\beta_L=0.02$, in place of 0.01, 
the freedom and lockdown periods increase by 3 and 5 days, respectively.
Finally, the bottom panel shows how the trajectory 
of the exposed $E$ and infected $I$ evolves in the two-dimensional plane (blue) and how it approaches the  
sliding region (red), then moving around it.\\
Fig. \ref{Fig3} shows the temporal lasting of the freedom (green) 
and lockdown (red) period as a function of the reproduction number 
(which was instead fixed to 1.3 to build Fig. \ref{Fig2}).
As expected, the freedom period required to control the epidemic
with a precision error similar to that displayed in the top panel of Fig. \ref{Fig2} decays as the reproduction number 
increases. Indeed, larger values of \eqref{RN} imply larger freedom or, also, a more aggressive pathogen agent.
The freedom period seems to decay exponentially, while the lockdown duration
slowly increases.\\
Fig. \ref{Fig3b} illustrates the robustness of our control law. 
We simulated different SEIR models by making parameters $\epsilon$
and $\gamma$ vary in the intervals already defined above.
During any simulated epidemic the proposed control law was applied and 
the percentage deviation
$$
100\% \frac{\max_t |I(t)-I_0|}{I_0}
$$
was computed after the initial transient. 
The histogram of all these values is visible
and one can see that the maximum percentage errors range from $12\%$ to $24\%$ 
(using the nominal
parameter values it turns out $20\%$, as seen before).
This well points out the effectiveness of our technique: without any knowledge on SEIR parameters, 
good control precision is always ensured.\\
All of these figures were then redone by using a more aggressive control law,
obtained by using $\lambda=0.6$.
Comparing the top panels of Fig. \ref{Fig2}  and Fig. \ref{Fig4},
one can see that the amplitude of the oscillation is now reduced.
The middle panel of Fig. \ref{Fig4} shows that the control is now obtained
with almost one month of freedom followed by one week of severe lockdown.
This is even more in line with the Australian approach.\\ 
Fig. \ref{Fig5} plots the temporal duration of the freedom (green) and lockdown (red) period, as a function of the reproduction number. Comments similar to those regarding Fig. \ref{Fig3} hold.
Finally, control robustness w.r.t. system parameters variations is documented
in Fig. \ref{Fig5b}. Now, the maximum percentage error is smaller than $14\%$,
hence ensuring high control precision.

\subsection{Effect of vaccines administration in improving the effectiveness of the control law}
We also investigated the combined effect of the administration of vaccines and of the control law
by simulating scenarios where a different number of people is vaccinated every day.
For this purpose, one can just subtract and add the positive input $V(t)$ 
to the equations governing $S(t)$ and $R(t)$, respectively.  
We assume that the vaccine is administrated at $t=60$
and then becomes active in the population 
after two months, hence $V(t)\neq0$ only for $t\geq120$ \citep{VirusVar2021}.  
The left and right top panels of Fig. \ref{FigVAX} show the temporal evolution of the infected $I(t)$
for $\lambda=0.2$ and $\lambda=0.6$, respectively.
Three different rates of vaccination are assumed, where $0.03\%,0.08\%$ or $0.16\%$
of the population is vaccinated every day. In Italy the last scenario corresponds  to
vaccinate almost $100000$ people per day. Using the smallest rate,
one can see how the oscillations period 
tends to increase as time progresses,
making $I(t)$ staying longer below $I_0$. 
In the other two cases, beyond this, the epidemic then quickly disappears.
The left and right bottom panels of Fig. \ref{FigVAX} illustrate the effect of the vaccines on the 
freedom and lockdown periods for $\lambda=0.2$ (left) and $\lambda=0.6$
(right) with the vaccination rate set to $0.08\%$.

\subsection{Final remarks on obtained results with and without vaccines}

\begin{figure*}
\center
{\includegraphics[scale=0.4]{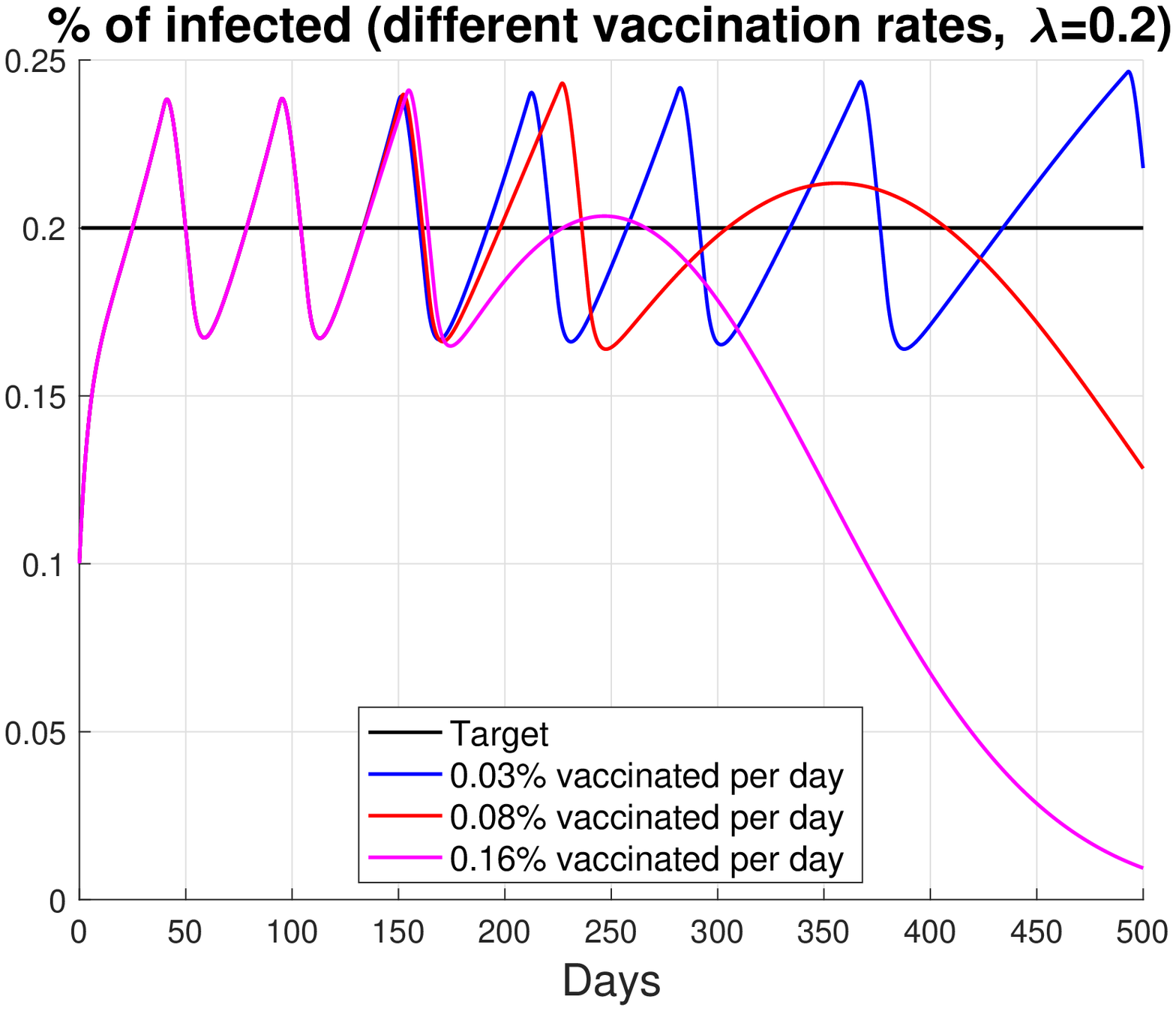}} \  {\includegraphics[scale=0.4]{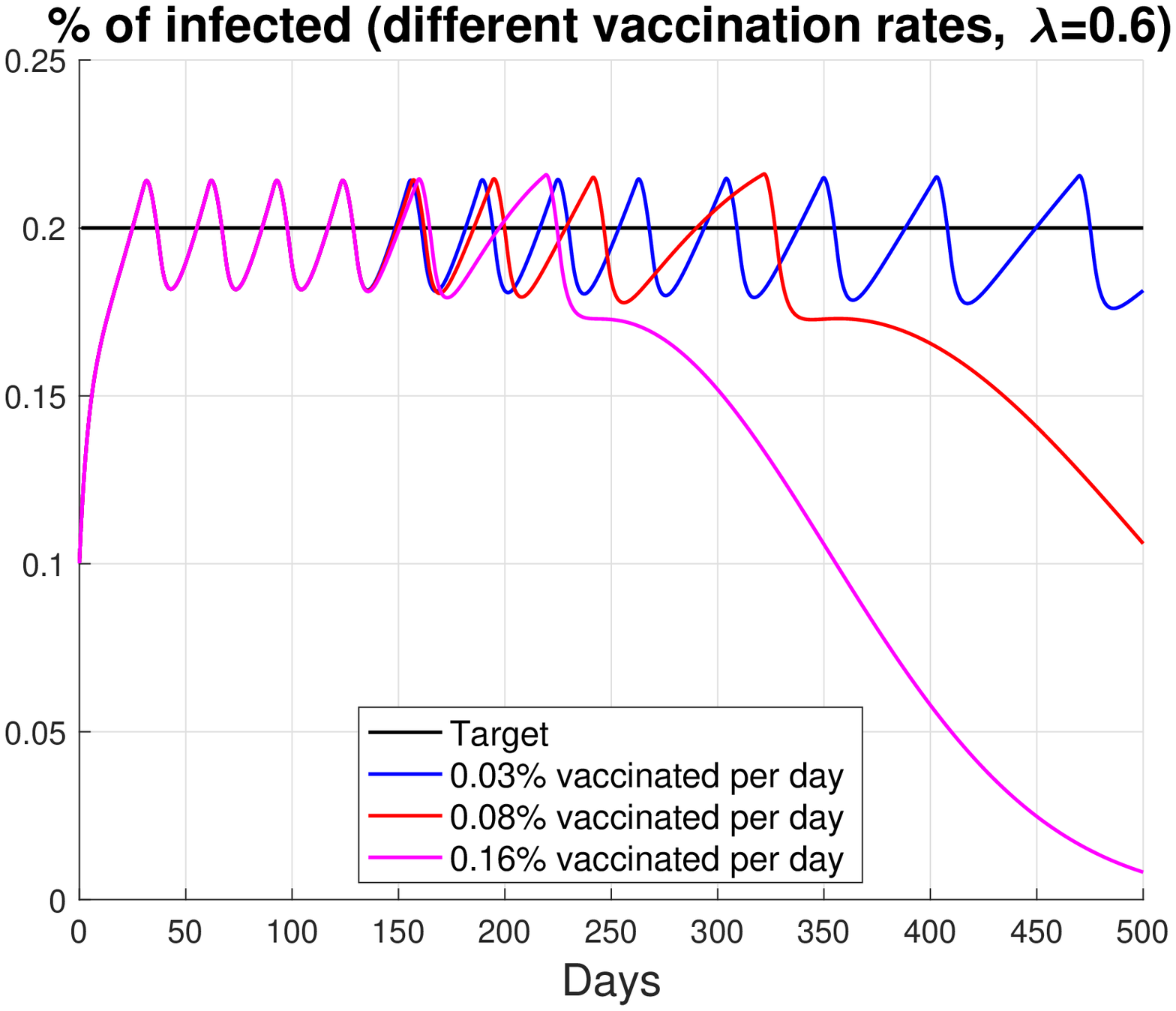}} \\
{\includegraphics[scale=0.4]{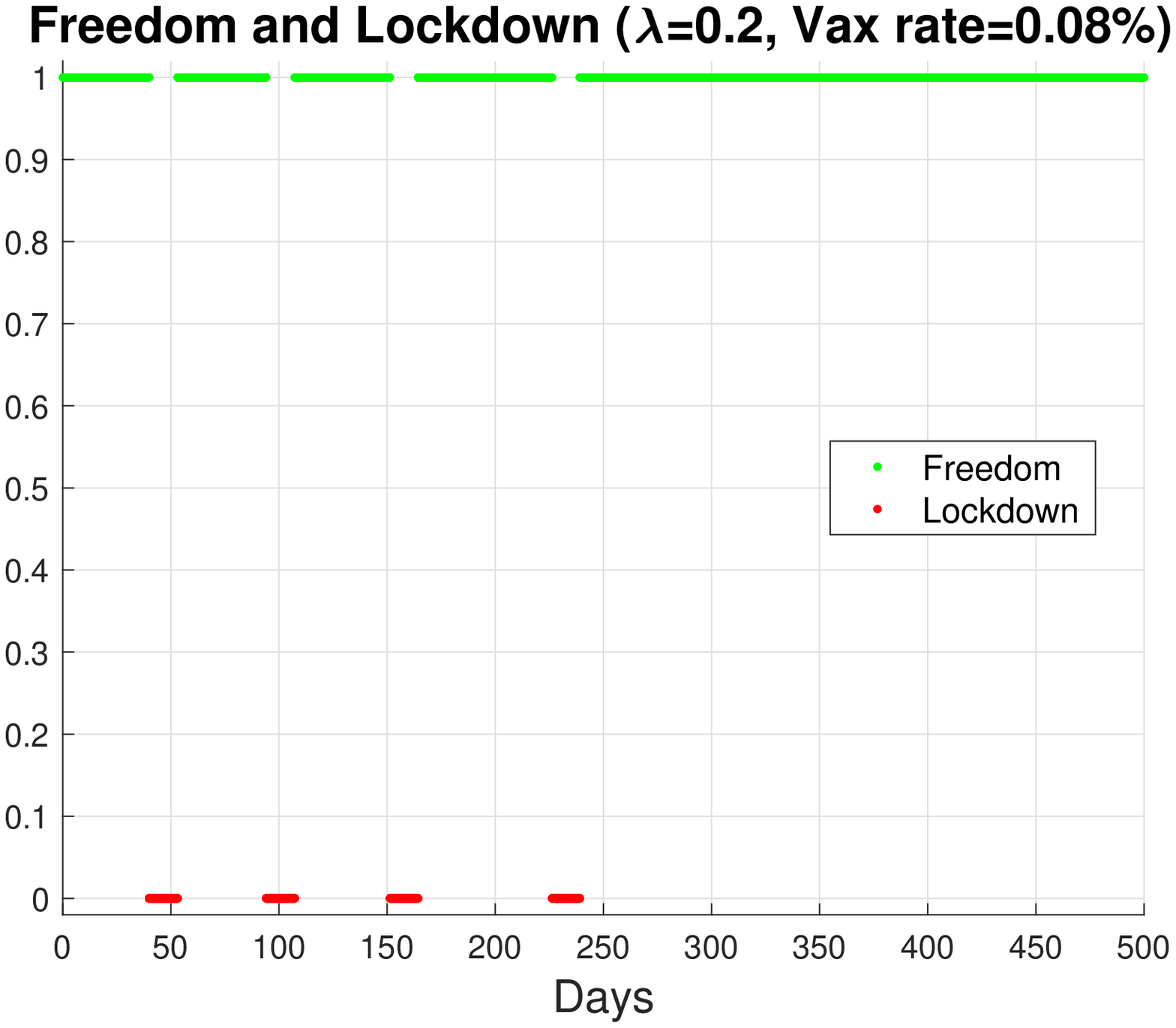}} \  {\includegraphics[scale=0.4]{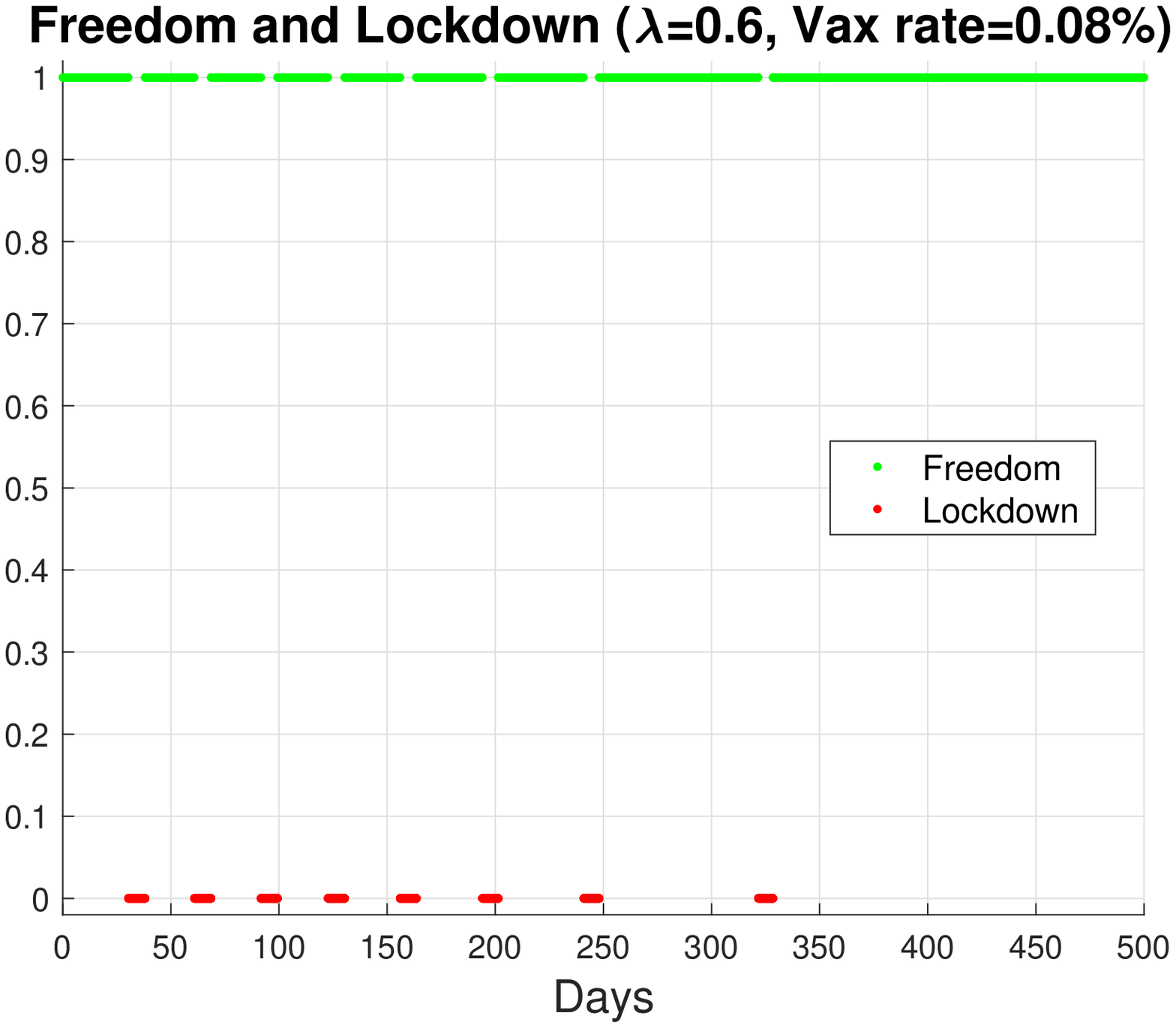}}
\caption{Use of vaccines administrated at $t=60$ and with effect starting after two months. \emph{Top} Percentages of infected and desired percentage (black) as a function of days for three different 
vaccination rates using $\lambda=0.2$ (left) or $\lambda=0.6$ (right). 
\emph{Bottom} Temporal evolution of freedom (green) and lockdown (red) periods
for $\lambda=0.2$ (left) and $\lambda=0.6$ (right)
with the vaccination rate set to $0.08\%$.}
\label{FigVAX}
\end{figure*}

We have seen in the top panels of Fig. \ref{FigVAX} that trajectories of the infected 
generated by our control law in presence of 
vaccines administration are aperiodic. Actually, this feature also holds in absence 
of vaccines: in the previous figures, e.g. top panel of Fig. \ref{Fig2}, 
they seem periodic only because of the limited time range there adopted. 
This is due to the following reasons:
\begin{itemize}
\item oscillations are the consequence of the use of non-ideal sliding mode connected with the chattering phenomenon 
induced by the choice $\phi>0$. 
Specifically, 
it induces an oscillation around the nominal trajectory on the sliding line 
which is moving linearly, and exponentially fast in time, along the $(E,I)$ plane towards the equilibrium point. 
Apart from the initial brief transient phase, the infected peak would be equal to $I_0$ in case of $\phi=0$, while in practice it turns out to be a little larger because of the chattering;
\item since $S(t)$ is monotone decreasing, the duty-cycle, defined as the ratio between the lockdown time and the total time of the oscillations, is continuously decreasing (this is visible only after a certain number of oscillations). 
This happens because, as time increases, the effective reproduction number,
that corresponds to the product between the basic reproduction number and $S(t)$ \citep{Fine1993}, 
is slowly decreasing
both during the freedom and lockdown period. Our simulations also show that the freedom phase is more sensitive 
to variations of $S(t)$, hence the period of the oscillations increases while the duty-cycle decreases as time progresses;
\item the lockdown threshold is not reached forever. In fact, when $S(t)$ becomes small enough, 
the epidemic dies down on its own, without the need of further lockdowns. Hence, the equilibrium point cannot be any longer maintained, the trajectory of infected leaves the sliding line and tends to zero (epidemic extinction).
\end{itemize}
Regarding the last point mentioned above,
the fact that the control is not maintained forever 
has also consequences on the study of epidemic final size \citep{DiLauro2021,Morris2021,Bliman2021}.
In particular, the control law will be disabled after the so-called herd immunity will take place, i.e. 
when the effective reproduction number during the freedom period is equal to one \citep{Fine1993}.
Specifically, this will happen at the time-instant $t_{herd}$ such that  $S(t_{herd})=:S_{herd}=\frac{\gamma}{\beta_F}$.
The fact that attainment of the herd immunity is eventually guaranteed by the proposed control law also implies that the arising of another epidemic wave is prevented.  
The exact final value $S(\infty)$ will then depend on the social behaviour after $t_{herd}$, but it will likely be a little smaller than $S_{herd}$ since the epidemic
will be going towards extinction. \\ 
Note also that in its ideal version (i.e. with no chattering), and neglecting the transient phase, the control law here described minimizes the epidemic duration
subject to the constraint $I(t) \leq I_0$.
In fact, if $I(t)=I_0$ for any $t$, one has $R(t)=\gamma\int_0^t \ I(\tau)d\tau=\gamma I_0 t$, so the epidemic length is given by $T=\frac{1-S_{herd}}{\gamma I_0}$.
It is easily seen that any other choice of $I(t)$, subject to the constraint $I(t) \le I_0$, leads to a longer epidemic duration.


  \section*{Conclusions}
 
Differently from the strategies adopted in USA and Europe, the Australian model
tends to suddenly set short-term severe lockdowns as soon as the number of infected exceeds a certain threshold. We have shown that this strategy can be described in mathematical terms  providing 
a connection with sliding-mode, an important branch of nonlinear control.
From this work, one cannot argue that the Australian approach is the best one. However, this paper develops some mathematical and simulation tools that strengthen it, showing that it can likely be a valid alternative solution for other countries. Our analysis also generalizes and improves it since any value of infected $I_0$ to be controlled 
can be chosen.
Our study 
permits also to simulate interesting scenarios of action which can also include the effect of vaccines administration.
It leads to a 
robust control law, having high-performance also in presence of scarce knowledge 
on the parameters
which regulate SARS-CoV-2 dynamics. When a certain threshold related to the number of either infected or people in intensive care is overcome, the epidemic can be promptly controlled. In fact, 
from any initial condition the control law permits to come back to the desired equilibrium point 
with an exponential speed.  
Such a control strategy reveals advantages in terms of sociability: also in other countries people could prefer to undergo a severe but 
short-term lockdown (e.g. lasting just one or two weeks) to then enjoy a much larger period of freedom (e.g. of one or two months).

\section{Appendix}

  \subsection{SEIR control}\label{Sec1A}

\noindent We report the theoretical analysis concerning
the SEIR while the SAIR and SEAIR cases are discussed in the last part of Appendix.\\
In \eqref{SEIReq}
we are interested in establishing the (possible) existence of non-zero equilibrium points for the only variables $I(t), E(t)$.
Let the equilibrium values be $I_0$ and $E_0$, respectively. From $\dot{I}(t)=0$, it holds that $\epsilon E_0=\gamma I_0$, while $\dot{E}(t)=0$ leads to  $\beta(t)S(t)=\gamma$. These equations admit at least one positive solution 
if and only if
\begin{equation}
\beta(t)=\beta_0(t):=\frac{\gamma}{S(t)}.
\label{1}
\end{equation}
Letting $W(t):=E(t)+I(t)$,
one has $\dot{W}(t)=[\beta(t)S(t)-\gamma]I(t)$. Recall that only two values are allowed for $\beta(t)$: $\beta_L$ (lockdown) and $\beta_F$ (freedom), so in case of a full freedom, the epidemic spread requires $\beta_F S(t)>\gamma$, while the lockdown is successful only if $\gamma>\beta_L S(t)$, so we need to assume
\begin{equation}\label{betaF}
\beta_F> \frac{\gamma}{S(t)}=\beta_0(t)>\beta_L>0.
\end{equation}
Therefore the positive equilibrium point exists for $\beta(t)=\beta_0(t)$, and we are only interested in situations in which $\beta_F>\beta_0(t)>\beta_L>0$. Let's briefly discuss the case where such inequality is not satisfied. If $\beta_0(t)<\beta_L$ the lockdown is not successful and we simply cannot face the epidemic spread, but only slow it. If $\beta_0(t)>\beta_F$, there is no need to set a lockdown, as the epidemic will disappear on its own. Under the assumption (\ref{betaF}) we can thus define $\delta(t)$ through
\begin{equation}
\beta(t)S(t)=\gamma+\delta(t)
\label{2}
\end{equation}
with $\delta(t)$ having opposite signs in the freedom/lockdown cases. By rewriting (\ref{SEIReq}) in terms of $\delta(t)$ and $W(t)$,
one has
$$
\begin{pmatrix}\dot{E}(t)=\gamma I(t)-\epsilon E(t)+\delta(t)I(t) \cr \dot{I}(t)=\epsilon E(t)-\gamma I(t) \end{pmatrix}
$$
that implies 
$$
\begin{pmatrix} \dot{W}(t)=\delta(t)I(t) \cr \dot{I}(t)=\epsilon W(t)-(\gamma+\epsilon)I(t)\end{pmatrix}
$$
with  $\epsilon W_0=(\gamma+\epsilon)I_0$.
Then, exploiting the following change of variables
$$
u(t):=\delta(t)I(t), \ x_1(t):=I(t)-I_0, \ x_2(t):=W(t)-W_0, 
$$
$$
{\bf x}(t)=\begin{pmatrix}x_1(t) & x_2(t)\end{pmatrix}^T
$$
one obtains the simple linear system 
\begin{equation}
\dot{\bf x}=\begin{pmatrix}\dot{x}_1 \cr \dot{x}_2\end{pmatrix}=\begin{pmatrix}-(\gamma+\epsilon) & \epsilon \cr 0 & 0\end{pmatrix}
\begin{pmatrix} x_1 \cr x_2\end{pmatrix}+\begin{pmatrix} 0 \cr 1\end{pmatrix} u:=F{\bf x}+gu.
\label{EQ}
\end{equation}



\noindent Some results from Sliding Modes Control Theory are now briefly recalled  
and then also specialized to the two-dimensional case.
\medskip

\noindent {\bf Proposition 1 [Equivalent input on a sliding surface].} \ Let $K{\bf x}=\begin{pmatrix}-h & -1\end{pmatrix}{\bf x}=0$ (which is equivalent to $hx_1+x_2=0$) be the desired sliding surface. The equivalent input $u_{eq}(t)$ which allows to remain in the sliding surface is obtained by imposing $0=K\dot{x}=KF{\bf x}+Kgu_{eq}=KF{\bf x}-u_{eq}$ and is therefore given by
\begin{equation}
u_{eq}=KF{\bf x}=h(\gamma+\epsilon)x_1-h\epsilon x_2
\label{3}
\end{equation}
and this leads to
$$
\begin{pmatrix} \dot{x}_1 \cr \dot{x}_2 \end{pmatrix}=\begin{pmatrix} -(\gamma+\epsilon) & \epsilon \cr h(\gamma+\epsilon) & -h\epsilon \end{pmatrix} \begin{pmatrix} x_1 \cr x_2 \end{pmatrix}, \ \dot{\bf x}:=A{\bf x}, \  \mbox{with} \ K\dot{\bf x}=KA{\bf x}=0.
$$ 
\hfill$\blacksquare$

\noindent By defining
\begin{equation}
\lambda:=\gamma+\epsilon(1+h)>0 \ \Leftrightarrow \ h=\frac{\lambda-(\gamma+\epsilon)}{\epsilon}
\label{LAM}
\end{equation}
the eigenvalues of $A$ are given by $-\lambda,0$. The negative eigenvalue ensures asymptotic stability, and this requires
$$
h \in \left(-\frac{\gamma+\epsilon}{\epsilon}, \ +\infty\right).
$$
In other words, $K\dot{\bf x}=0$ implies $K{\bf x}(t)=0$ for any $t \ge t_1$ if $K{\bf x}(t_1)=0$, and this implies $x_2(t)=-hx_1(t)$ for any $t \ge t_1$, so that after some easy computations
\begin{equation}
\dot{x}_2=-\lambda x_2
\label{EXP}
\end{equation}
which makes the dynamics on the sliding line expressed by a linear system with eigenvalue $-\lambda<0$. So, once $\lambda>0$ has been chosen, it suffices to evaluate $h$ through (\ref{LAM}) in order to obtain the expression $x_2+hx_1=0$ of the sliding line and to ensure that the dynamics of ${\bf x}(t)$ on the sliding line only depends on $\lambda$, and is therefore completely independent of the system parameters $\gamma, \epsilon$.
\medskip

\noindent For what regards the control law, following Sliding Modes Theory, a simple possibility is the discontinuous law
\begin{equation}
u(t)=\alpha(t,\text{sign}[K{\bf x}]) \text{sign}[K{\bf x}], \ \alpha(t,\text{sign}[K{\bf x}]) \ge  \alpha_0>0
\label{CL}
\end{equation}
where $\alpha(\cdot,\cdot)$ is any uniformly lower-bounded (by some positive constant $\alpha_0>0$) function, and where the (possible) dependence of $\alpha$ on both $t$ and $\text{sign}[K{\bf x}]$ (because its value could be different over and under the sliding line) have been put in evidence. For simplicity sake, in the following we will omit the arguments of $\alpha$, when not strictly necessary. Now, we need to investigate when this law is able to reach (and then to
maintain) the state trajectory on the sliding surface.
\medskip

\noindent {\bf Proposition 2 \ [Sufficient condition for the (LOCAL) sliding establishment].} \ Let ${\mathcal L}_0$ be a subset (possibly either only a point or the whole line) of the (stabilizing) sliding line ${\mathcal L}$ containing the equilibrium point 
${\bf x}_{\bf eq}=0$. Assume that a closed neighborhood ${\mathcal I}$ of ${\mathcal L}_0$ exists such that 
\begin{itemize}
\item ${\mathcal I}$ is an invariant region (which means ${\bf x}(0) \in {\mathcal I}$ implies ${\bf x}(t) \in {\mathcal I}, \ \forall t \ge 0$)
\item for any $x \in {\mathcal I}/{\mathcal L}$ it holds  $\alpha_0>c+\text{sign}[K({\bf x}-{\bf x_{eq})}]KF{\bf x}$, for some $c>0$
\end{itemize}
Then the control law (\ref{CL}) permits to reach ${\mathcal L}_0$ in finite time, starting from any initial condition $x(0) \in {\mathcal I}$  and then the trajectory never escapes from ${\mathcal L}_0$.\\ 
\noindent {\bf Proof:} \ From (\ref{EQ}) it easily follows (in this case ${\bf x}_{\bf eq}=0$)
$$
[K\dot{\bf x}]=KF{\bf x}-\alpha \text{sign}[K{\bf x}] \ \Rightarrow \ \text{sign}[K{\bf x}][K\dot{\bf x}]=\text{sign}[K{\bf x}]KF{\bf x}-\alpha<-c  
$$
which is surely verified if
$$
\alpha_0>c+\text{sign}[K{\bf x}]KF{\bf x}
$$
and easily implies $K{\bf x}(t_1)=0$ for some $0 \le t_1 \le \frac{|KF{\bf x}(0)|}{c}$. From $t=t_1$ onwards, $K {\bf x}(t)$ cannot become different from zero, otherwise a contradiction would arise as a consequence of the opposite signs of $K{\bf x}(t)$ and of its derivative over and under ${\mathcal L} \cap {\mathcal I}$. The invariance property is required both for the inequality validity until the sliding line is reached, and, together with the monotone (exponential) movement on ${\mathcal L}$ towards ${\bf x}_{\bf eq}=0$ (see (\ref{EXP})), for avoiding to exit from ${\mathcal L}$ along the sliding line, once ${\mathcal L}$ is reached. \hfill$\blacksquare$
\medskip

The following corollary is then immediately obtained.\\

\noindent {\bf Corollary 1 \ [Sufficient condition for the (GLOBAL) sliding establishment].} \ Let ${\mathcal I}$ satisfy the assumptions of Proposition 2, and assume that for any $x(0) \in {\mathbb R}_+^2$ the corresponding trajectory reaches ${\mathcal I}$ in finite time. Then the sliding establishment is globally guaranteed.
\medskip


\noindent The following proposition is our main result. It shows when the control law (\ref{CL}) satisfies the conditions reported above for either local or global sliding. The proof is quite technical and shows how (\ref{CL}) and the sliding line introduced above coincide, respectively, with the control law exposed before (\ref{SlidingSurface}) and with (\ref{SlidingSurface}) itself.\\

\noindent {\bf Proposition 3 [Effectiveness of the control law (\ref{CL})].} For $\lambda>0$ small enough, the control law (\ref{CL}) satisfies the conditions for the global sliding establishment. Furthermore, the same control law satisfies the conditions for the local sliding establishment for any choice of $\lambda$ and of the SEIR parameters. 

\noindent {\bf Proof:} \  First, recall the definition of the scalar $\lambda$, i.e.
$$
\lambda=\gamma+\epsilon(1+h)>0.
$$

\noindent Now, we want to prove that the control law (\ref{CL}) satisfies the conditions expressed in Proposition 2 for any $\lambda>0$ and the conditions expressed in Corollary 1 if $\lambda$ is small enough. The proof is split into some steps.
\medskip

\noindent {\bf Step 1 \ [Computation of $K{\bf x}$, $KF{\bf x}$, $\alpha(t,\text{sign}[K{\bf x}])$ and $u(t)$].} \ First of all, $x_2$ can be expressed in terms of $x_1, \dot{x}_1$. In fact
$$
\dot{I}=\epsilon W-(\gamma+\epsilon)I \ \Rightarrow \ \dot{x}_1=\dot{I}=\epsilon (x_2+W_0)-(\gamma+\epsilon)(I_0+x_1)
$$
$$
=\epsilon x_2 -(\gamma+\epsilon)x_1+[\epsilon W_0-(\gamma+\epsilon)I_0]=\epsilon x_2 -(\gamma+\epsilon)x_1
$$
from which
$$
x_2=\frac{1}{\epsilon}[\dot{x}_1+(\gamma+\epsilon)x_1] \ \Rightarrow \ K{\bf x}=-[hx_1+x_2]=-\frac{1}{\epsilon}\{[(h+1)\epsilon+\gamma]x_1+\dot{x}_1\}
$$
$$
=-\frac{1}{\epsilon}[\dot{x}_1+\lambda x_1]=-\frac{1}{\epsilon}[\dot{I}+\lambda(I-I_0)].
$$
Recalling also $u(t)=\delta(t)I(t)$ and (\ref{2}), it holds that $u(t)=[\beta(t)S(t)-\gamma]I(t)$, which leads, in case of freedom and lockdown respectively, to
$$\footnotesize
\begin{array}{lcl} 
\alpha&=&\alpha_F(t):=\alpha(t,1)=[\beta_FS(t)-\gamma]I(t)=\dot{W}(t)=\dot{E}(t)+\dot{I}(t) \ \mbox{necessarily if} \ K{\bf x}>0, \ \mbox{from} \ \alpha>0, \cr
\alpha&=&\alpha_L(t):=\alpha(t,-1)=[\gamma-\beta_LS(t)]I(t)=-\dot{W}(t)=-\dot{E}(t)-\dot{I}(t) \ \mbox{necessarily if} \ K{\bf x}<0, \ \mbox{from} \ \alpha>0.
\end{array}
$$
Therefore the control law
\begin{equation}
u(t)=\alpha(t,\text{sign}[K{\bf x}]) \text{sign}[K{\bf x}]=-\alpha(t)[\dot{I}(t)+\lambda(I(t)-I_0)]
\label{LAW}
\end{equation}
easily follows, together with the sliding line expression
\begin{equation}
K{\bf x}=0 \Leftrightarrow \ \dot{I}+\lambda(I-I_0)=\epsilon E+(\lambda-\gamma)I-\lambda I_0=0
\label{SLIN}
\end{equation}
and with the expression of the sign of $K{\bf x}$
$$
\text{sign}[K{\bf x}]=-\text{sign}[\dot{I}+\lambda (I-I_0)]=-\text{sign}[\epsilon E+(\lambda-\gamma)I-\lambda I_0].
$$
The last expression shows why no measurements on $E$ are needed: all the information about $E$ is included in the knowledge of the pair $(I,\dot{I})$. Above, $-\lambda<0$ is exactly the eigenvalue we want to obtain, whose choice is completely independent of the knowledge of $\epsilon, \gamma$. Finally, recalling the expression of $KF{\bf x}$, it holds that
$$
KF{\bf x}=h(\gamma+\epsilon)x_1-h\epsilon x_2=h(\gamma+\epsilon)x_1-h[\dot{x}_1+(\gamma+\epsilon)x_1]
$$
$$
=-h\dot{x}_1=-h\dot{I}=h[\gamma I -\epsilon E]=\frac{\lambda-(\gamma+\epsilon)}{\epsilon}[\gamma I -\epsilon E].
$$

\noindent {\bf Step 2 \ [The conditions in Proposition 2 (in case of $\lambda<\gamma+\epsilon$)].} \ The conditions to be checked become
$$
\begin{array}{lcl}
K{\bf x}&<&0 \ \Leftrightarrow \ E>\frac{\gamma-\lambda}{\epsilon}I+\frac{\lambda}{\epsilon}I_0 \ \Rightarrow \ \alpha>c+\frac{(\gamma+\epsilon)-\lambda}{\epsilon}[\epsilon E-\gamma I], \cr
K{\bf x}&>&0 \ \Leftrightarrow \ E<\frac{\gamma-\lambda}{\epsilon}I+\frac{\lambda}{\epsilon}I_0 \ \Rightarrow \ \alpha>c+\frac{(\gamma+\epsilon)-\lambda}{\epsilon}[\gamma I-\epsilon E]
\end{array}
$$
which, recalling the expression of $\alpha$ reported in the previous step, allows to express the previous implications as follows
$$
\begin{array}{lcl}
E&>&\frac{\gamma-\lambda}{\epsilon}I+\frac{\lambda}{\epsilon}I_0 \ \Rightarrow \ [\gamma-\beta_LS(t)]I>c+\frac{(\gamma+\epsilon)-\lambda}{\epsilon}[\epsilon E-\gamma I], \cr
E&<&\frac{\gamma-\lambda}{\epsilon}I+\frac{\lambda}{\epsilon}I_0 \ \Rightarrow \ [\beta_FS(t)-\gamma]I>c+\frac{(\gamma+\epsilon)-\lambda}{\epsilon}[\gamma I-\epsilon E].
\end{array}
$$
Now assume $\lambda<\gamma+\epsilon$. After some computations we obtain, equivalently
$$
\begin{array}{lcl}
E&>&\frac{\gamma-\lambda}{\epsilon}I+\frac{\lambda}{\epsilon}I_0 \ \Rightarrow \ E<\left[\frac{\gamma}{\epsilon}+\frac{\gamma-\beta_LS(t)}{(\gamma+\epsilon)-\lambda}\right]I-\frac{c}{(\gamma+\epsilon)-\lambda}, \cr
E&<&\frac{\gamma-\lambda}{\epsilon}I+\frac{\lambda}{\epsilon}I_0 \ \Rightarrow \ E>\left[\frac{\gamma}{\epsilon}-\frac{\beta_FS(t)-\gamma}{(\gamma+\epsilon)-\lambda}\right]I+\frac{c}{(\gamma+\epsilon)-\lambda}.
\end{array}
$$

\noindent {\bf Step 3 \ [Computation of the region ${\mathcal I}$ satisfying the conditions in Proposition 2 (in case of $\lambda<\gamma+\epsilon$)}]. \ In view of the previous results, we need to consider three straight lines of the form $E=m_iI+q_i$, with
$$
m_1=\frac{\gamma-\lambda}{\epsilon}I, \ q_1=\frac{\lambda}{\epsilon}I_0, \ m_2=\left[\frac{\gamma}{\epsilon}+\frac{\gamma-\beta_LS(t)}{(\gamma+\epsilon)-\lambda}\right]I, \ q_2=-\frac{c}{(\gamma+\epsilon)-\lambda}, \
$$
$$
m_3=\left[\frac{\gamma}{\epsilon}-\frac{\beta_FS(t)-\gamma}{(\gamma+\epsilon)-\lambda}\right]I, \ q_3=\frac{c}{(\gamma+\epsilon)-\lambda}.
$$
Consider also other two lines defined by
$$
m_4=\frac{\gamma}{\epsilon}, \ q_4=0, \ m_5=max \ [ \ 0,m_3 \ ], \ q_5=q_3
$$
where the first one expresses all the (possible) equilibrium points, while the second one is needed since $m_3$ could be negative. It is easy to see that, for $\lambda>0$ and $c>0$ small enough, a point $I_1=I_1(c,\lambda,I_0)>0$ (continuously depending on $\lambda, c, I_0$, and tending to the origin as $\lambda$ and $c$ go to zero) exists such that
$$
I \ge I_1 \ \Rightarrow \ m_2I+q_2 \ge m_1I+q_1, \ m_2I+q_2 > m_4I+q_4
$$
while it holds
$$
m_4I+q_4\ge m_1I+q_1 \ \mbox{if and only if} \ I \ge I_0.
$$
Hence, the first implication ($E>m_1I+q_1$ implies $E<m_2I+q_2$) is verified in the closed region
$$
{\mathcal R}_1(I_3):=\{ \ (I,E)>0: \  m_2I+q_2\ge E\ge m_1I+q_1, \ I \ge I_3 > I_1 \ \}
$$
which has a piecewise-linear boundary. Since $E > m_4I+q_4$ implies both that $I$ is increasing and $I+E$ is decreasing, this prevents any trajectory to cross either the line $I=I_3$ or the line $E=m_2I+q_2$, for any $I_3$ such that $I_1<I_3<I_0$.
\medskip

\noindent Analogously, for $\lambda>0$ and $c>0$ small enough, also a point $I_2=I_2(c,\lambda,I_0)>0$ (with the same properties $I_1$ has) exists such that
$$
I \ge I_2 \ \Rightarrow \ m_1I+q_1 \ge m_5I+q_5, \ m_4I+q_4 \ge m_5I+q_5
$$
and the second implication ($E<m_1I+q_1$ implies $E>m_3I+q_3$) is verified in the closed region
$$
{\mathcal R}_2(I_3):=\{ \ (I,E)>0: \  m_1I+q_1\ge E \ge m_5I+q_5, \ I+E \ge (1+m_4)I_3 \ \}
$$
which has a piecewise-linear boundary. Since $E < m_4I+q_4$ implies both that $I$ is decreasing and $I+E$ is increasing, this prevents any trajectory to cross either the line $I+E=I_3(1+m_4)$ or the line $E=m_5I+q_5$, for any $I_3$ such that $I_3>I_2$. So it holds $E\ge m_5I+q_5\ge m_3I+q_3$ a fortiori in the whole region itself.
\medskip

\noindent Now it suffices to choose $I_3 > max(I_1,I_2)$ and $I_3<I_0$ to obtain the region 
$\mathcal{I}={\mathcal R}_1(I_3) \cup {\mathcal R}_2(I_3)$ as a closed neighborhood of almost all the sliding line satisfying the conditions of Proposition 2. Recall that $I_1, I_2$ continuously depend on $c,\lambda,I_0$ and that $I_1=I_2=0$ for $\lambda=c=0$. Then, choosing $c,\lambda$ small enough, we can obtain $I_3$ arbitrarily small. 
In other words, at least an interval $0<\lambda<\lambda_0$ exists such that the assumptions of Proposition 2 are satisfied in the aforementioned region, which contains almost all the sliding line except for an arbitrarily small segment. 
\medskip

\noindent {\bf Step 4 \ [Global convergence to the sliding line if $\lambda>0$ is small enough].}
\medskip

\noindent The proof that, for any $(I(0),E(0))>0$, the couple $(I(t),E(t))$ will eventually reach the aforementioned region $\mathcal{I}$ in finite time (hence ensuring the sliding functioning after a finite time for any initial condition, 
at least for $\lambda>0$ small enough) is simple but tedious. It is therefore omitted. 
\medskip

Below, we now verify that, independently of the system parameters  and of $\lambda>0$, both the conditions in Proposition 2 are always satisfied. Such a local property is needed for allowing the arising of sliding modes, even though it cannot guarantee it in a global sense, i.e. starting from any initial condition.
\medskip

\noindent {\bf Step 5 \ [Local (at least) convergence to the sliding line for any $\lambda>0$].}
\medskip

\noindent Starting from $K{\bf x}=-hx_1-x_2$, we can express $x_2=-hx_1-K{\bf x}$, and resorting to a change of variables $(x_1,x_2) \leftrightarrow (x_1,x_3)$, with $x_3:=K{\bf x}$, the differential equations can be rewritten as follows
$$
\dot{x}_1=-\epsilon\left[\frac{\lambda}{\epsilon}x_1+x_3\right], \ \text{sign}[x_3]\dot{x}_3=\text{sign}[x_3][\lambda-(\gamma+\epsilon)]\left[\frac{\lambda}{\epsilon}x_1+x_3\right]-\alpha(t)
$$
Recalling $\alpha(t)=\delta(t)I(t)=\delta(t)[I_0+x_1]$ (see Step 1), where $\alpha$ depends on both $t$ (through $S(t)$) and the lockdown/freedom state, let's assume $|x_1|\le\frac{I_0}{2}$ and denote with $\alpha_0$ the minimum value $\delta(t)\frac{I_0}{2}$ can assume. Choose now any $\mu>0$ such that
$$
|\lambda-(\gamma+\epsilon)|\mu<\frac{\alpha_0\epsilon}{2\lambda} \ \mbox{and} \ \mu<\frac{I_0}{2}
$$
and consider the closed neighborhood of ${\bf x}_{\bf eq}=0$ given by
$$
{\mathcal I}:=\{ \ (x_1,x_3): \ |x_1| \le \mu, \ |x_3| \le \frac{\lambda}{\epsilon}\mu \ \}.
$$
From
$$
|\lambda-(\gamma+\epsilon)| \cdot \left|\frac{\lambda}{\epsilon}x_1+x_3\right|\le|\lambda-(\gamma+\epsilon)| \cdot \left[\frac{\lambda}{\epsilon}|x_1|+|x_3|\right]
$$
$$
\le\frac{2\lambda}{\epsilon}|\lambda-(\gamma+\epsilon)|\mu<\alpha_0:=(1-\nu)\alpha_0, \ 0<\nu<1
$$
it follows $\text{sign}[x_3]\dot{x}_3<-\nu\alpha_0$, so that ${\bf x}(0) \in {\mathcal I}$ implies that $x_3(t)$ monotonically converges to zero in finite time, if $|x_1(t)|\le \mu$ forever. But this easily follows from $\dot{x}_1=-\epsilon\left[\frac{\lambda}{\epsilon}x_1+x_3\right]$, by analyzing the evolution of $x_1(t)$ which moves towards the straight-line $x_3=-\frac{\lambda}{\epsilon}x_1$. This does not allow ${\bf x}(t)$ to escape from ${\mathcal I}$, hence showing that both the conditions in Proposition 2 are always satisfied.
\hfill$\blacksquare$
\medskip



%

Some remarks are now in order.

\begin{itemize} 
\item It has been shown that the sliding line attracts all the trajectories in finite time for $\lambda$ belonging to a suitable (small) interval. However, this doesn't prevent the possibility of obtaining a sliding functioning also for other (possibly all) values of $\lambda>0$. In 
fact, Proposition 3 states only a \emph{sufficient} condition. 
Note however that we have not only characterized some situations in which the sliding-mode technique can be applied, but also, through 
Proposition 3, that any choice of $\lambda$ and of the SEIR parameters is compatible with the existence of a sliding-mode evolution.
\item Once the sliding line is reached at some point $(I_1,E_1)$, two scenarios are possible. Either the trajectory crosses the sliding line (in case the point $(I_1,E_1)$ doesn't belong to the attractive neighborhood ${\mathcal I}$ of the equilibrium point $(I_0,E_0)$ - see Proposition 2), or it remains on the sliding line forever (actually until $S(t)$ reaches the value $S_{herd}$ and the epidemic is quickly going towards its extinction). In the latter case, $I(t)$ evolves accordingly to the exponential monotone behaviour described by $I(t)=I_0+(I_1-I_0)e^{-\lambda (t-t_1)}$, where $t_1$ satisfies $I(t_1)=I_1$ (see (\ref{EXP})). Such a behaviour becomes therefore completely independent of the (possibly unknown) values of the SEIR parameters, from $t_1$ onwards. What depends instead on $(\gamma,\epsilon)$ is the sliding line expression and, in particular, the value of $E_0$. This latter is however of no interest since the target is to control the steady-state value $I_0$. The parameters values could instead influence both the transient phase and the establishment of the sliding-mode, but our simulations show that it is obtained in all the cases of interest. It is worth also remarking that reaching the sliding line doesn't guarantee the sliding-mode establishment. In fact, usually only an attractive neighborhood ${\mathcal I}$ is available 
and some line crossings can occur before the sliding mode takes place also using perfect sliding,
i.e. avoiding chattering by setting $\phi=0$. This phenomenon is also graphically illustrated
in Fig. \ref{FigAppRev}.
\begin{figure*}
\center
{\includegraphics[scale=0.3]{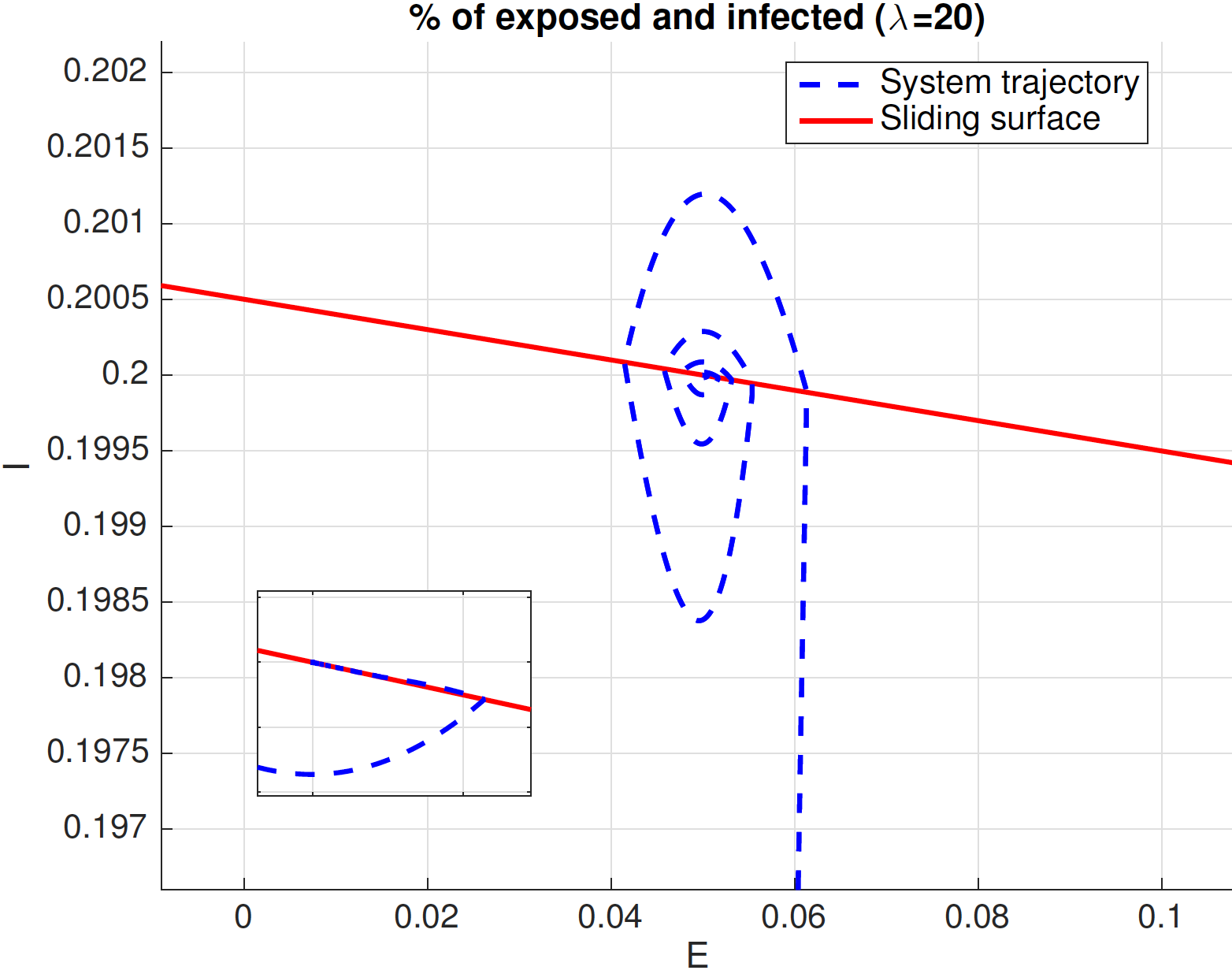}}
\caption{State trajectories on the $(E,I)$-plane
with same setting used to build Fig. \ref{Fig2} except that 
$\phi$ is set to zero (no chattering phenomenon can happen) and $\lambda=20$.
The smaller part of the figure provides a zoom showing how the trajectory eventually enters the sliding line
and then remains over it, after some previous sliding line crossings.
}
\label{FigAppRev}
\end{figure*}
\item From a practical viewpoint, one needs to alternate lockdown and freedom periods, and this leads to an oscillation around $I_0$. This can be obtained by modifying the control law as follows: if $K{\bf x}>s$ then $u(t)=\alpha(t,1)$, if $K{\bf x}<-s$ then $u(t)=-\alpha(t,-1)$, otherwise $u(t)$ remains locked at the previous value. Here, $s$ represents a threshold that can be suitably tuned by using the parameter $\phi$ entering the control law. This means that we can  modulate the amplitude (and the duration) of the oscillations to define lockdown/freedom periods of practical meaning, without significantly affecting either the mean value (that remains almost equal to $I_0$) or the duty-cycle of the ON/OFF signal associated with the lockdown presence or absence. This is the well-known \emph{chattering} phenomenon, which consists of finite-frequency oscillations around the sliding surface (the theoretical frequency in the ideal case would be infinite). An illustration is given in Fig. \ref{FigMM} where the effect of two different values of $\phi$ on the 
control of the infected is illustrated.

\begin{figure*}
\center
{\includegraphics[scale=0.34]{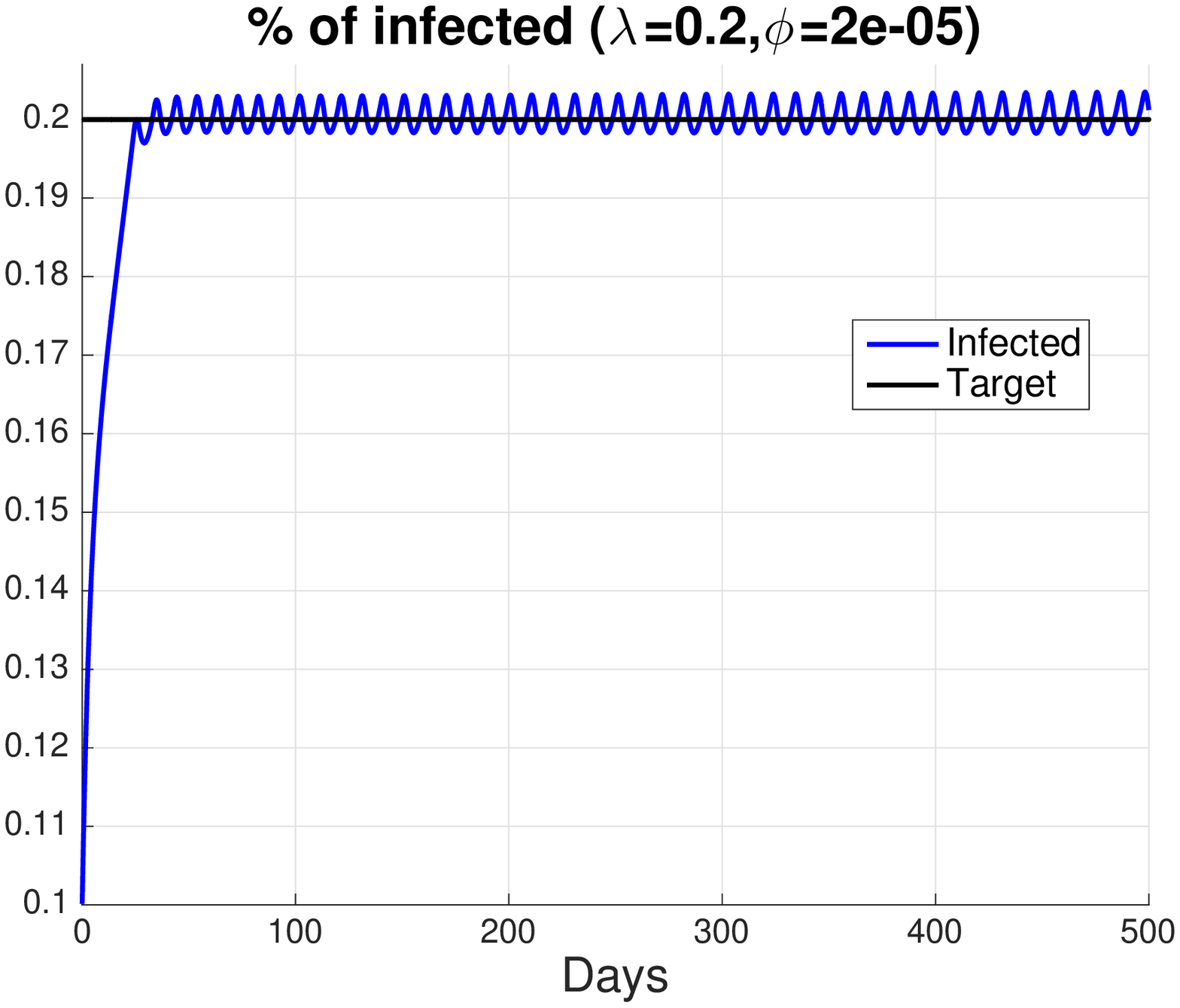}} \  {\includegraphics[scale=0.34]{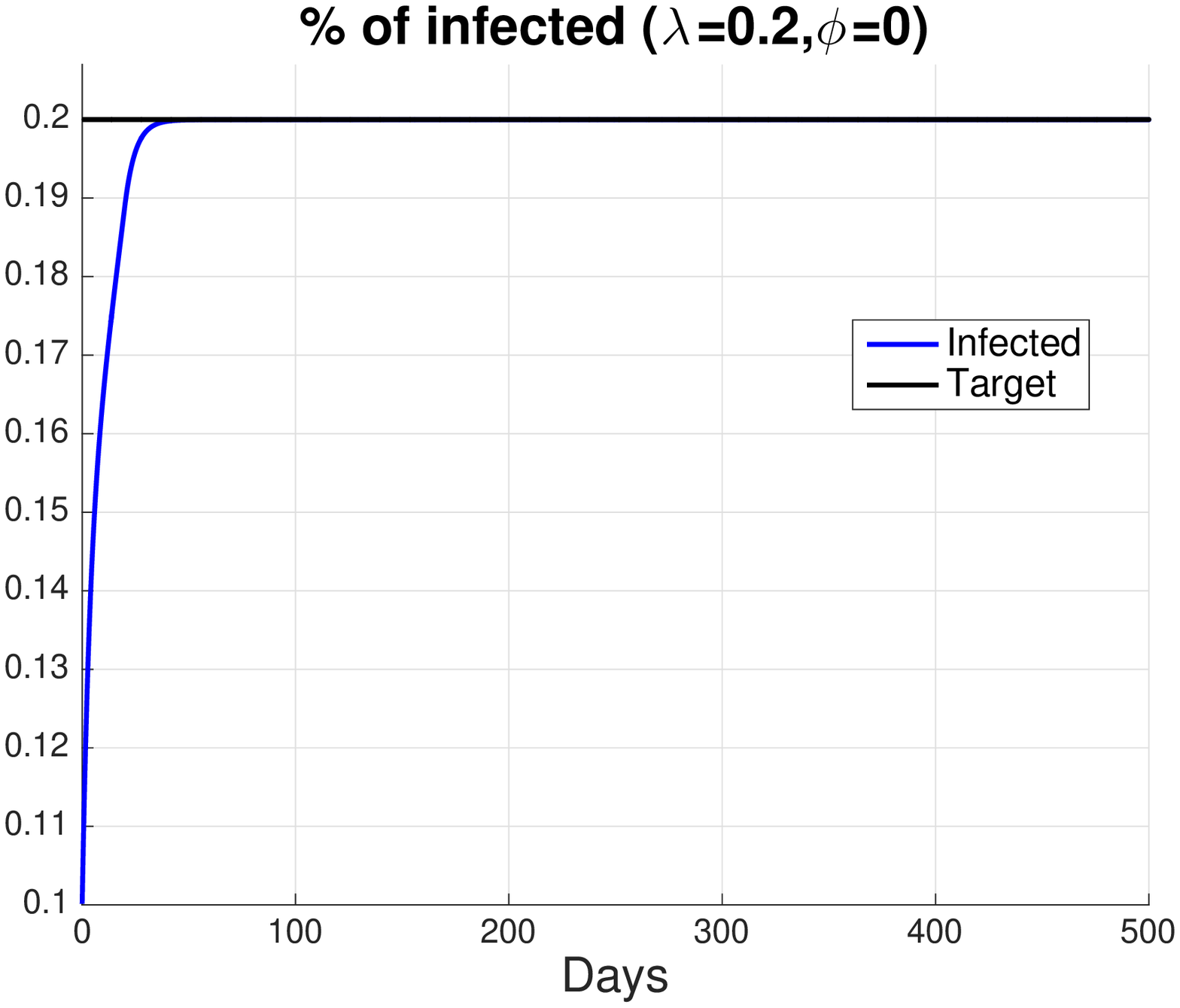}}
\caption{Percentage of infected (blue) and desired percentage (black) as a function of days.
The same setting used to build the top panel of Fig. \ref{Fig2} is adopted except that the parameter
$\phi$ entering the control law is set to smaller values.
}
\label{FigMM}
\end{figure*}

\item Once the steady-state values are reached, the system behaves like if $\beta(t)$ assumes exactly the (time-varying) value $\beta_0(t)$ given by (\ref{1}). In other words,  the structure of the control law guarantees in an automatic way that the reproduction number becomes exactly equal to $1$. 
\item The condition $\beta_FS(t)>\gamma$ cannot hold forever since $S(t)$ is strictly decreasing.  At the epidemic beginning $S(t) \simeq 1$, so $W(t)$ satisfies $\dot{W}(t)\simeq[\beta(t)-\gamma]I(t)>0$ if $\beta(t)=\beta_F$. But $\dot{W}(t)>0$ holds true until $S(t)$ becomes small enough to be no longer able to feed $E(t), I(t)$ in such a way that the equilibrium point is maintained. In fact, $S(t)$ is monotone and strictly decreasing, so if $\beta_FS(t_1)=\gamma$ for some $t_1>0$ (actually $t_1=t_{herd}$ and $S(t_1)=S_{herd}$, as already previously discussed), the $W(t)-$evolution in case of full freedom satisfies
\begin{equation}
\dot{W}(t)=[\beta_FS(t)-\gamma]I(t)\le 0
\label{W}
\end{equation}
which leads to the epidemic extinction without further control (lockdowns) actions.
This means that at a certain time instant the evolution will escape from the sliding line. This is not a problem: it corresponds to epidemic extinction and control laws are no longer required. 
As $S(t)$ becomes smaller, the freedom period becomes longer and the duty-cycle tends to privilege freedom to lockdown as time 
advances. Then, at a certain temporal instant the lockdown disappears and the trajectory of infected people leaves the steady-state value when the epidemic is going towards extinction. Examples are the red and magenta lines in the top panels of Fig. \ref{FigVAX} and the bottom panels in the same figure
where these
phenomena are accelerated by vaccines administration that makes $S(t)$ decrease faster. 

\end{itemize}

\subsection{SAIR and SEAIR control}

\subsubsection{SAIR model} \ \noindent Now, consider the SAIR model given by
 \begin{subequations}\label{SAIReq}
\begin{align}
\dot{S}(t)&=- \beta(t)S(t)\big(A(t)+I(t)\big) \\
\dot{A}(t)&= \beta(t)S(t)\big(A(t)+I(t)\big) - \big(\epsilon_1+\epsilon_2\big)A(t) \\
\dot{I}(t)&= \epsilon_1A(t)-\gamma I(t) \\ 
\dot{R}(t)&=\epsilon_2A(t)+\gamma I(t) 
\end{align}
\end{subequations}

\begin{figure*}
\center
{\includegraphics[scale=0.38]{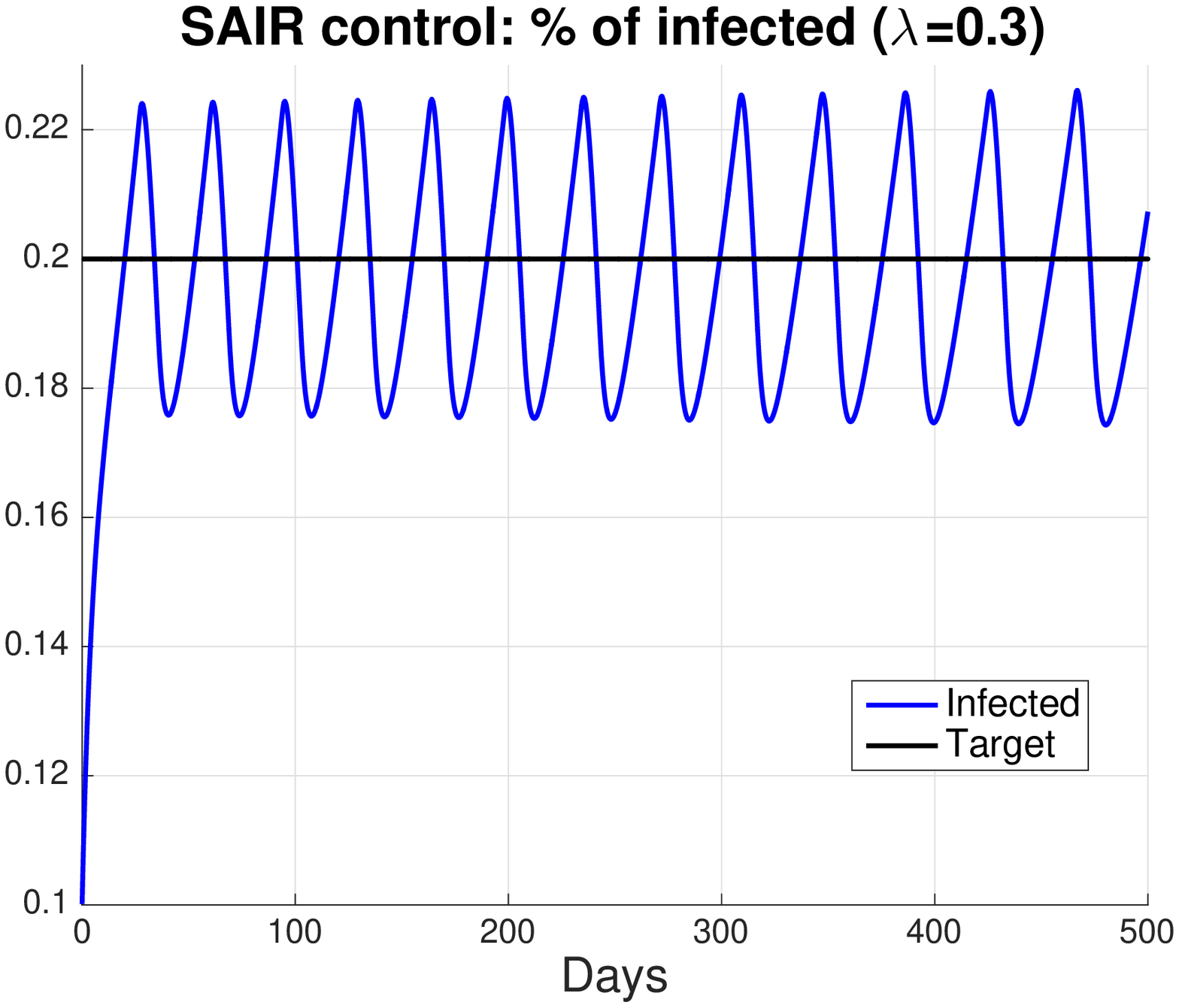}} \ \vspace{0.2cm}
{\includegraphics[scale=0.38]{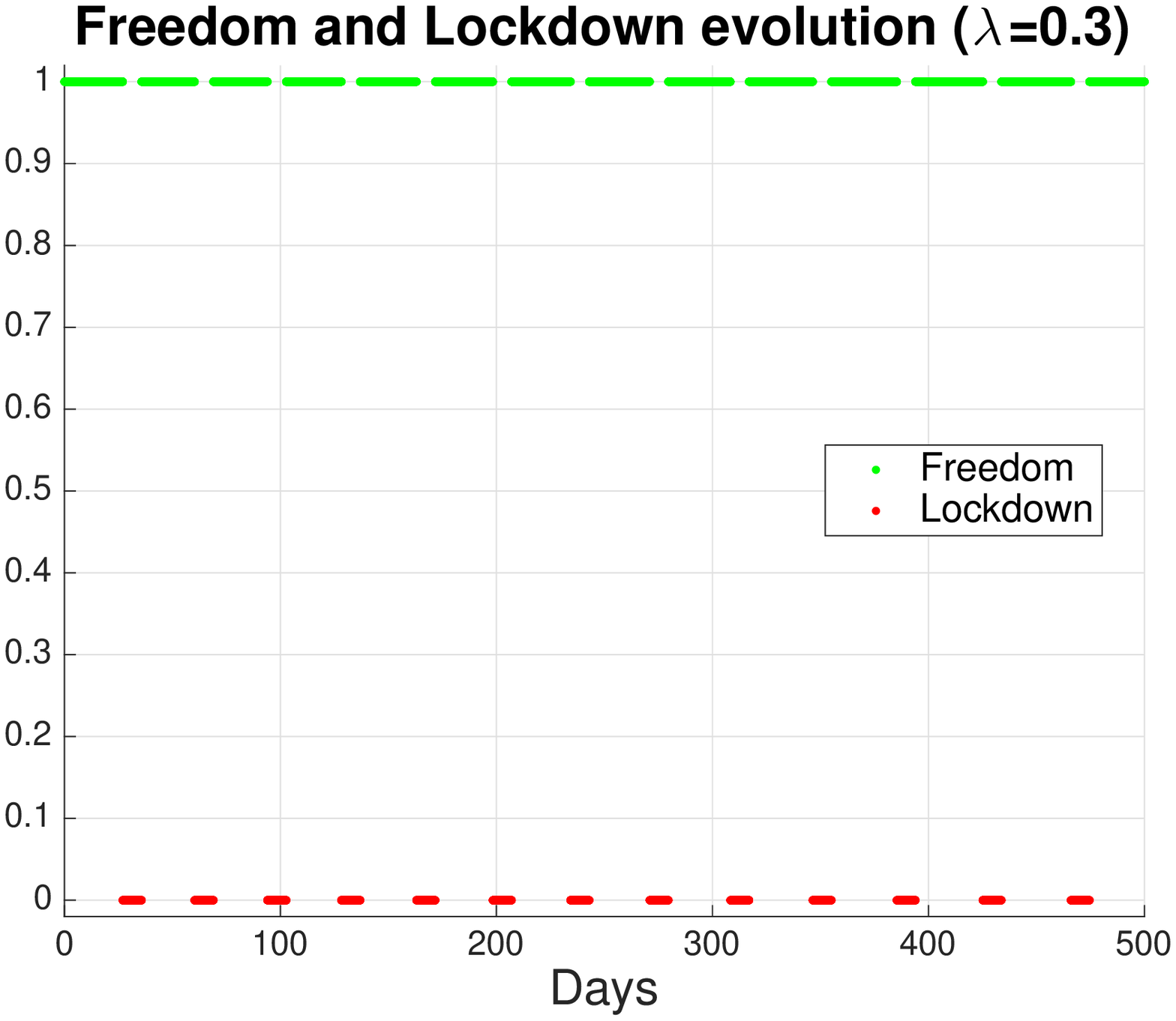}} \\ \vspace{0.2cm}
{\includegraphics[scale=0.38]{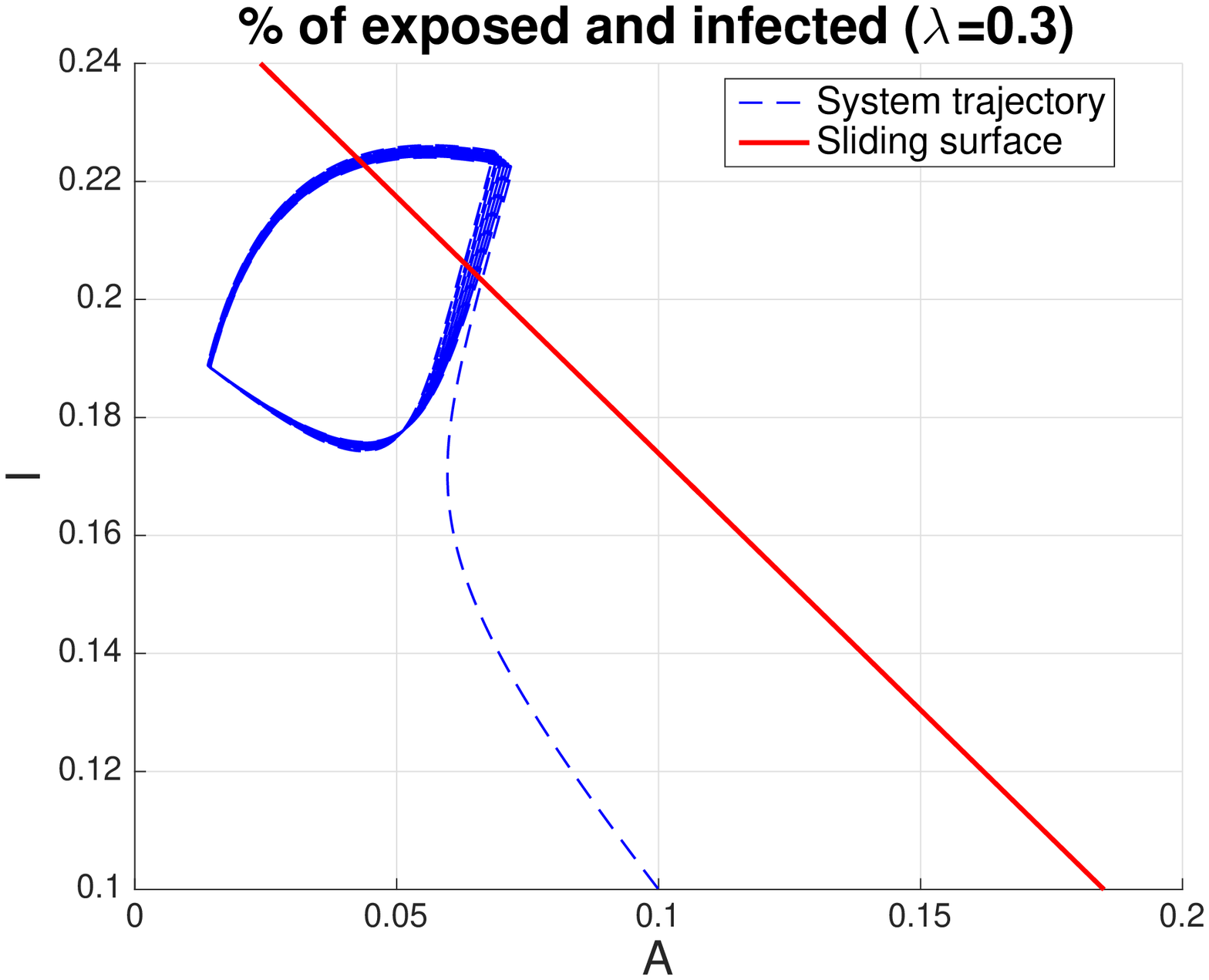}}
\caption{Control of SAIR model. Results returned by the nonlinear control law with $\lambda=0.3$. \emph{Top left} Percentage of infected (blue) and desired percentage (black) as a function of days.
\emph{Top right} Temporal evolution of freedom (green) and lockdown (red) periods. 
After a transient, almost one month of freedom and one week of severe lockdown are alternated.
\emph{Bottom} Trajectory describing how the percentage of exposed and infected varies in time (blue) and sliding surface (red).}
\label{FigSI1}
\end{figure*}

Similarly to what done in the SEIR case, first we focus only on the equations related to the evolutions of $A,I$ to
determine the structure of the possible non-zero equilibrium points. The conditions for their existence
are
$$
A_0=\frac{\gamma}{\epsilon_1}I_0, \ \beta(t)S(t)(A_0+I_0)=(\epsilon_1+\epsilon_2)A_0
$$
$$
\Downarrow
$$
$$
\beta(t)S(t)=\frac{\gamma(\epsilon_1+\epsilon_2)}{\gamma+\epsilon_1}.
$$
Hence, we can define
$$
\beta(t)S(t)=\frac{\gamma(\epsilon_1+\epsilon_2)}{\gamma+\epsilon_1}+\delta(t)
$$
which requires
$$
\beta_F>\beta_0(t)>\beta_L, \ \text{where now} \ \beta_0(t):=\frac{\gamma(\epsilon_1+\epsilon_2)}{(\gamma+\epsilon_1)S(t)}
$$
in order to obtain $\delta(t)$ having opposite signs in case of freedom and lockdown
(at least until $S(t)$ is large enough).  
Simple calculations provide
$$
\begin{pmatrix}\dot{I} \cr \dot{A} \end{pmatrix}=\begin{pmatrix}-\gamma & \epsilon_1 \cr \frac{\gamma(\epsilon_1+\epsilon_2)}{\gamma+\epsilon_1} & -\frac{\epsilon_1(\epsilon_1+\epsilon_2)}{\gamma+\epsilon_1} \end{pmatrix}\begin{pmatrix} I \cr A \end{pmatrix}+\begin{pmatrix}0 \cr 1\end{pmatrix} u, 
$$
$$
u(t)=[A(t)+I(t)]\delta(t)
$$
which, by introducing nominal values and a matrix $K$ defining our sliding surface, allows to obtain
$$
I=I_0+x_1, \ A=A_0+x_2, \ K=\begin{pmatrix}-h & -1\end{pmatrix} 
$$
$$
\Downarrow
$$
$$
\dot{\bf x}=\begin{pmatrix}-\gamma & \epsilon_1 \cr \frac{\gamma(\epsilon_1+\epsilon_2)}{\gamma+\epsilon_1} & -\frac{\epsilon_1(\epsilon_1+\epsilon_2)}{\gamma+\epsilon_1}\end{pmatrix}{\bf x}+\begin{pmatrix}0 \cr 1\end{pmatrix}u.
$$
The equivalent input  is given by
$$
u_{eq}=KF{\bf x}=\begin{pmatrix}\gamma h-\frac{\gamma(\epsilon_1+\epsilon_2)}{\gamma+\epsilon_1} & -\epsilon_1h+\frac{\epsilon_1(\epsilon_1+\epsilon_2)}{\gamma+\epsilon_1}\end{pmatrix}{\bf x}
$$
and leads to the following autonomous linear system
$$
\dot{\bf x}:=P{\bf x}, \ P=\begin{pmatrix}-\gamma & \epsilon_1 \cr \gamma h & -\epsilon_1h\end{pmatrix}
$$
whose eigenvalues are $0,-\lambda$.
In particular, $\lambda>0$ is given by
$$
\lambda=\gamma+\epsilon_1h \ \Leftrightarrow \ h=\frac{\lambda-\gamma}{\epsilon_1}
$$
and makes 
$$
K{\bf x}=-hx_1-x_2=\frac{\gamma-\lambda}{\epsilon_1}I-A+\frac{\lambda}{\epsilon_1}I_0
$$
$$
=\frac{\gamma-\lambda}{\epsilon_1}I-\frac{\dot{I}+\gamma I}{\epsilon_1}+\frac{\lambda}{\epsilon_1}I_0=-\frac{1}{\epsilon_1}[\lambda (I-I_0)+\dot{I}].
$$
We thus obtain
$$
[\beta(t)S(t)-\frac{\gamma(\epsilon_1+\epsilon_2)}{\gamma+\epsilon_1}][A(t)+I(t)]=\delta(t)[A(t)+I(t)]=
$$
$$
=u(t)=\alpha(t,\text{sign}[K{\bf x}])\text{sign}[K{\bf x}]=
$$
$$
=-\alpha(t,\text{sign}[K{\bf x}])\text{sign}[\lambda (I-I_0)+\dot{I}], \ \alpha(t,\text{sign}[K{\bf x}]) \ge \alpha_0>0
$$
which implies the same control law obtained in the SEIR case,
with the constant $\phi \ge 0$ playing exactly the same role previously discussed, i.e.
$$
[\lambda (I-I_0)+\dot{I}]>\phi \ \Rightarrow \ \mbox{LOCKDOWN}, 
$$
$$
[\lambda (I-I_0)+\dot{I}]<-\phi \ \Rightarrow \ \mbox{FREEDOM}.
$$
The only thing that changes is the relationship between the eigenvalue $-\lambda$, the model parameters and the constant $h$. Proposition 3 is satisfied in this case too, with arguments omitted due to space constraints and because they are quite similar to those developed in the SEIR case.\\
Now, we report a simulation concerning SAIR control
with system parameters all set to values well supported in the literature \citep{Crisanti2020,Gatto2020,Giordano2020,Incubation}.
We set $\epsilon_1=0.2$ (on average asymptomatic become
symptomatic in 5 days),  $\epsilon_2=0.07$ (asymptomatic heal in two weeks)
and $\gamma=0.05$ (symptomatic heal in 20 days). The same reproduction numbers
adopted in the SEIR case then define the contact rate during the freedom and lockdown periods.
As documented in Fig. \ref{FigSI1},
the choice $\lambda=0.3,\phi=10^{-4}$ allows to control SARS-CoV-2 epidemic alternating
almost 32 days of freedom and 8 days of lockdown.

\subsubsection{SEAIR model} 
\begin{figure*}
\center
{\includegraphics[scale=0.38]{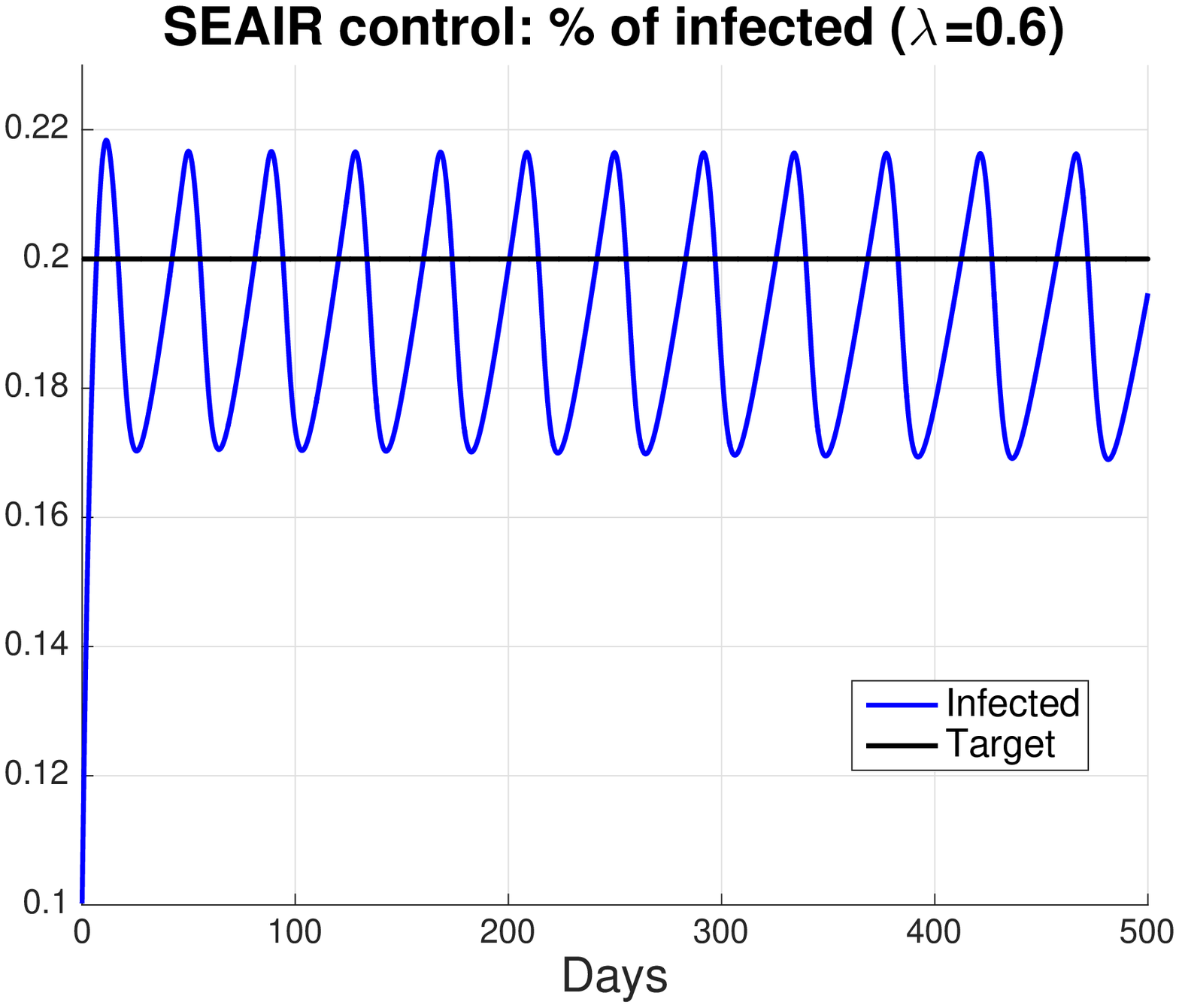}} \ 
{\includegraphics[scale=0.38]{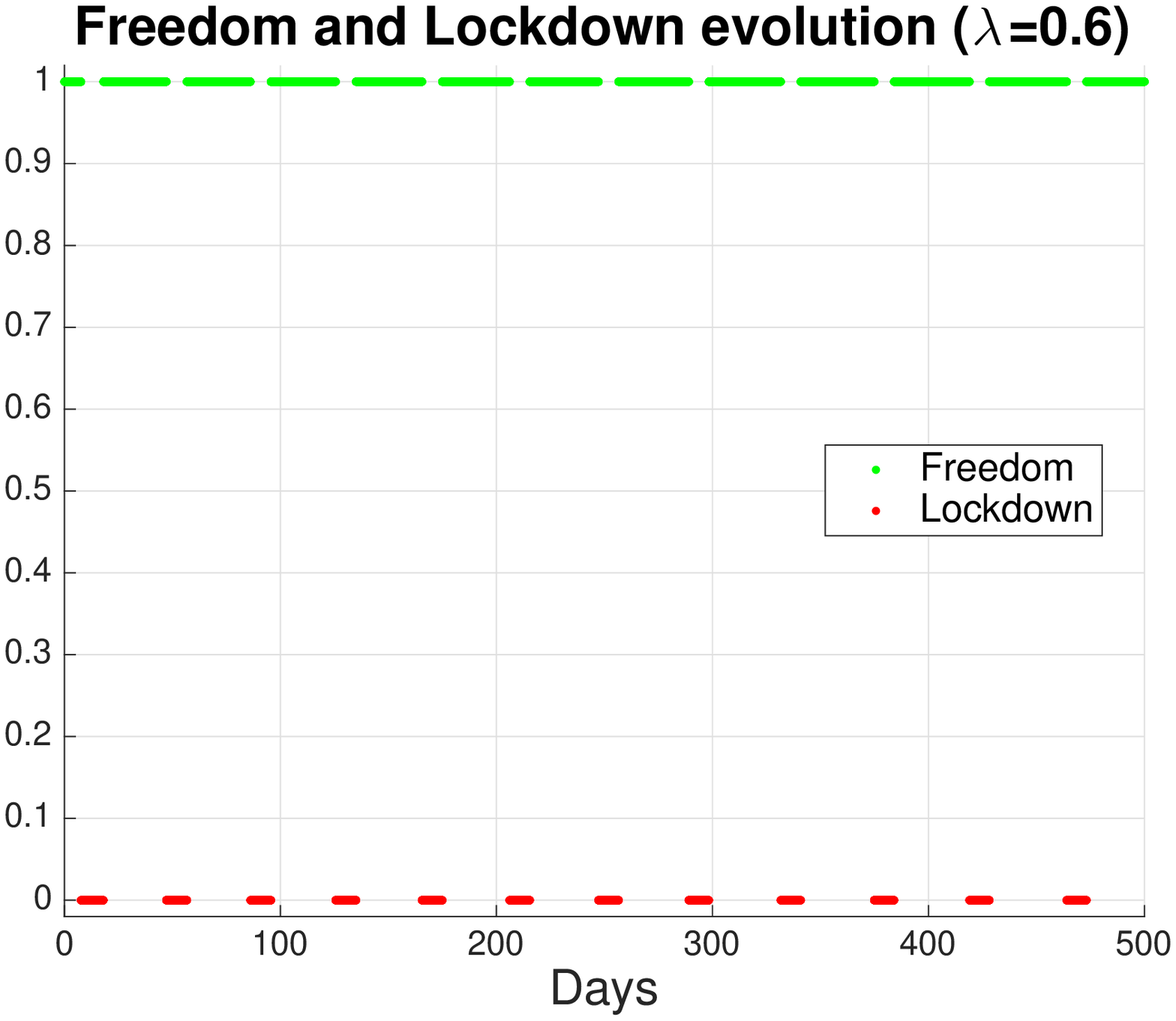}} 
\caption{Control of SEAIR model. Results returned by the nonlinear control law with $\lambda=0.6$  and $\mu=0$. \emph{Left} Percentage of infected (blue) and desired percentage (black) as a function of days.
\emph{Right} Temporal evolution of freedom (green) and lockdown (red) periods. 
After a transient, almost one month of freedom and one week of severe lockdown are alternated.}
\label{FigSI2}
\end{figure*}

The SEAIR model is a generalization of the SEIR and SAIR 
that contains both the exposed and the asymptomatic class. It is given by
 \begin{subequations}
\begin{align}
\dot{S}&=-\beta(t)S(t)(A+I)\\ 
\dot{E}&=\beta(t)S(t)(A+I)-\epsilon E\\
\dot{A}&=\epsilon E-(\epsilon_1+\epsilon_2)A\\
\dot{I}&=\epsilon_1A-\gamma I\\
 \dot{R}&=\epsilon_2A+\gamma I.
\label{SEAIR}
\end{align}
\end{subequations}
In this case, we have to focus on three differential equations governing $E,A,I$ in order to investigate non-zero equilibrium points:
$$
E_0=\frac{\gamma(\epsilon_1+\epsilon_2)}{\epsilon_1\epsilon}I_0, \ A_0=\frac{\gamma}{\epsilon_1}I_0, \ \beta(t)S(t)=\frac{\gamma(\epsilon_1+\epsilon_2)}{\gamma+\epsilon_1}.
$$
Now, we define $\delta(t)$ through
$$
\beta(t)S(t)=\frac{\gamma(\epsilon_1+\epsilon_2)}{\gamma+\epsilon_1}+\delta(t)
$$
which requires
$$
\beta_F>\beta_0(t)>\beta_L, \ \text{with once again} \ \beta_0(t):=\frac{\gamma(\epsilon_1+\epsilon_2)}{(\gamma+\epsilon_1)S(t)}
$$
to obtain $\delta(t)$ having opposite signs. Simple calculations lead to
$$
\begin{pmatrix}\dot{I} \cr \dot{A} \cr \dot{E}\end{pmatrix}=\begin{pmatrix}-\gamma & \epsilon_1 & 0 \cr 0 & -(\epsilon_1+\epsilon_2) & \epsilon \cr \frac{\gamma(\epsilon_1+\epsilon_2)}{\gamma+\epsilon_1} & \frac{\gamma(\epsilon_1+\epsilon_2)}{\gamma+\epsilon_1} & -\epsilon\end{pmatrix}\begin{pmatrix}I \cr A \cr E\end{pmatrix}+\begin{pmatrix}0 \cr 0 \cr 1\end{pmatrix}u, 
$$
$$
u(t)=[A(t)+I(t)]\delta(t).
$$
In this case, two coefficients $h,k$ are needed in the matrix $K$, i.e. 
$$
I=I_0+x_1, \ A=A_0+x_2, \ E=E_0+x_3, \ K=\begin{pmatrix}-h & -k & -1\end{pmatrix} 
$$
and this implies
$$
\dot{\bf x}=\begin{pmatrix}-\gamma & \epsilon_1 & 0 \cr 0 & -(\epsilon_1+\epsilon_2) & \epsilon \cr \frac{\gamma(\epsilon_1+\epsilon_2)}{\gamma+\epsilon_1} & \frac{\gamma(\epsilon_1+\epsilon_2)}{\gamma+\epsilon_1} & -\epsilon\end{pmatrix}{\bf x}+\begin{pmatrix}0 \cr 0 \cr 1\end{pmatrix}u.
$$
The equivalent input and the matrix $P$ are given by
$$
\begin{array}{lcl}
u_{eq}=KF{\bf x}=\begin{pmatrix}\gamma h-\frac{\gamma(\epsilon_1+\epsilon_2)}{\gamma+\epsilon_1} & -\epsilon_1h+k(\epsilon_1+\epsilon_2)-\frac{\gamma(\epsilon_1+\epsilon_2)}{\gamma+\epsilon_1} & -\epsilon k+\epsilon\end{pmatrix}{\bf x},
\end{array}
$$
hence
 $$
 \dot{\bf x}:=P{\bf x}
$$
with
$$
P=\begin{pmatrix}-\gamma & \epsilon_1 & 0 \cr 0 & -(\epsilon_1+\epsilon_2) & \epsilon \cr \gamma h & -\epsilon_1h+k(\epsilon_1+\epsilon_2) & -\epsilon k\end{pmatrix}.
$$
The characteristic polynomial of $P$ is
$$
\begin{array}{lcl}
\Delta(s):=s[s^2+s(\gamma+\epsilon_1+\epsilon_2+\epsilon k)+[\gamma(\epsilon_1+\epsilon_2)+\gamma\epsilon k+\epsilon_1\epsilon h]
\end{array}
$$
$$
\begin{array}{lcl}
=s[s^2+\Delta_1s+\Delta_0], \ \begin{pmatrix}h \cr k\end{pmatrix}=\frac{1}{\epsilon_1\epsilon}\begin{pmatrix}-\epsilon_1 & 1 \cr \epsilon_1 & 0\end{pmatrix}\begin{pmatrix}\Delta_1-(\gamma+\epsilon_2+\epsilon_1) \cr \Delta_0-\gamma(\epsilon_1+\epsilon_2)\end{pmatrix}
\end{array}
$$
which uniquely defines $h,k$ in terms of $\Delta_1,\Delta_0$ (one-to-one linear correspondence). 
Stability requires $\Delta_1,\Delta_0>0$.\\
Let us now express $K{\bf x}$ in terms of the only measurable variable $I$. 
To this aim, first we write $A$ by resorting to the equation of $\dot{I}$:
$$
A=\frac{1}{\epsilon_1}\dot{I}+\frac{\gamma}{\epsilon_1}I \ \Rightarrow \ \dot{A}=\frac{1}{\epsilon_1}\ddot{I}+\frac{\gamma}{\epsilon_1}\dot{I}.
$$
Then, exploiting also the equation of $\dot{A}$, one has 
$$
\frac{1}{\epsilon_1}\ddot{I}+\frac{\gamma}{\epsilon_1}\dot{I}=\dot{A}=\epsilon E-(\epsilon_1+\epsilon_2)A
$$
$$
\Downarrow
$$
 $$
 E=\frac{1}{\epsilon \epsilon_1}\ddot{I}+\frac{\gamma}{\epsilon \epsilon_1}\dot{I}+\frac{\epsilon_1+\epsilon_2}{\epsilon}A
$$
$$
=\frac{1}{\epsilon \epsilon_1}\ddot{I}+\frac{\gamma+\epsilon_1+\epsilon_2}{\epsilon \epsilon_1}\dot{I}+\frac{\gamma(\epsilon_1+\epsilon_2)}{\epsilon \epsilon_1}I 
$$
from which
$$
A=\frac{1}{\epsilon_1}\dot{I}+\frac{\gamma}{\epsilon_1}I, \ E=\frac{1}{\epsilon \epsilon_1}\ddot{I}+\frac{\gamma+\epsilon_1+\epsilon_2}{\epsilon \epsilon_1}\dot{I}+\frac{\gamma(\epsilon_1+\epsilon_2)}{\epsilon \epsilon_1}I. 
$$
This implies
$$
K{\bf x}=-h(I-I_0)-k(A-A_0)-(E-E_0)
$$
$$
=-\frac{1}{\epsilon \epsilon_1}\{[h\epsilon \epsilon_1+k\gamma\epsilon+\gamma(\epsilon_1+\epsilon_2)](I-I_0)+[k\epsilon+\gamma+\epsilon_1+\epsilon_2]\dot{I}+\ddot{I}\}.
$$
Recalling the expressions of $h,k,E_0,A_0$, this leads to the simple expression
$$
K{\bf x}=-\frac{1}{\epsilon_1\epsilon}[\Delta_0(I-I_0)+\Delta_1\dot{I}+\ddot{I}], \ \Delta_1, \Delta_0>0
$$
which represents the natural generalization of the SAIR and SEIR cases and 
defines the following (more sophisticated) control law
$$
\Delta_0(I-I_0)+\Delta_1\dot{I}+\ddot{I}>\phi \ \Rightarrow \ \mbox{LOCKDOWN}, 
$$
$$
\Delta_0(I-I_0)+\Delta_1\dot{I}+\ddot{I}<-\phi \ \Rightarrow \ \mbox{FREEDOM}
$$
with $\phi \ge 0, \Delta_1, \Delta_0>0$. The sliding surface
$$
K{\bf x}=0 \ \Leftrightarrow \ \Delta_0(I-I_0)+\Delta_1\dot{I}+\ddot{I}=0
$$
can be also formulated in terms of the state variables as follows
$$
(\gamma\Delta_1-\Delta_0-\gamma^2)I+\epsilon_1[(\gamma+\epsilon_1+\epsilon_2)-\Delta_1]A-\epsilon_1\epsilon E+\Delta_0I_0=0.
$$
One can notice the analogy with the previous control law and sliding surface expression, but the 
difference is that now the second order derivative of $I$ is also needed. 
This can make control harder since it could be difficult to obtain good estimates of $\ddot{I}$. 
However, using $f$ to indicate a generic positive scalar, one can see that
$$
\text{sign}[f\Delta_0(I-I_0)+f\Delta_1\dot{I}+\ddot{I}]=\text{sign}\left[\frac{\Delta_0}{\Delta_1}(I-I_0)+\dot{I}+\frac{1}{f\Delta_1}\ddot{I}\right]
$$
$$
:=\text{sign}[\lambda (I-I_0)+\dot{I}+\mu\ddot{I}]
$$
with $\lambda,\mu$ that now can be arbitrary positive real numbers.
As $\mu \rightarrow 0^+$, we obtain that $\ddot{I}$ disappears and the control law reduces to that seen before.  
Under a theoretical perspective, $\mu=0$ is not admissible  since no real sliding mode (the movement on the sliding surface after a transient phase) is possible. 
However, we are not interested in a pure sliding mode as we want to alternate periods of lockdown/freedom of significant duration. In this respect, while the presence of $\dot{I}$ is important since it contains the information about the trend,
choosing $\mu=0$ may be an admissible choice. In fact, from a practical viewpoint, if the aim is just to alternate periods of lockdown/freedom of significant duration, by resorting to a very small $\mu>0$ or to $\mu=0$ may lead to small differences in the control. Simulation results suggest that this happens.  In particular, we simulated SEAIR control
setting $\epsilon=0.3$ (exposed become asymptomatic in 3 days on average),
$\epsilon_1=0.3$ (asymptomatic become
symptomatic in 3 days),  $\epsilon_2=0.07$ (asymptomatic heal in two weeks)
and $\gamma=0.05$ (symptomatic heal in 20 days). The same reproduction numbers
adopted in the SEIR and SAIR case then define the contact rate during the freedom and lockdown periods.
Fig. \ref{FigSI2} shows the 
results obtained with $\lambda=0.6,,\phi=10^{-4},\mu=0$. SARS-CoV-2 is now controlled by alternating
almost 35 days of freedom and 9 days of lockdown.




\end{document}